\documentclass[
 preprint,
nofootinbib,
 amsmath,amssymb,
 aps,
]{revtex4}
\usepackage{graphicx}% Include figure files
\usepackage{dcolumn}% Align table columns on decimal point
\usepackage{bm}% bold math
\usepackage{graphicx}% Include figure files
\usepackage{dcolumn}% Align table columns on decimal point
\usepackage{bm}% bold math
\usepackage{hyperref}% add hypertext capabilities
\usepackage[normalem]{ulem}
\usepackage{color}
\usepackage{caption}
\usepackage{subcaption}
\usepackage{amsmath}
\usepackage{hyperref}

\begin{document}

\title{ Neutrino Mixing Phenomenology: \boldsymbol{$A_4$} Discrete Flavor Symmetry with Type-I Seesaw Mechanism}

\author{Animesh Barman}
    \thanks{Email address: animesh@tezu.ernet.in  (Corresponding author)}
\author{Ng. K. Francis}
    \thanks{Email address: francis@tezu.ernet.in}
    \author{Hrishi Bora}
    \thanks{Email address: hrishi@tezu.ernet.in}
    
\affiliation{Department of Physics, Tezpur University, Tezpur - 784028, India}


\begin{abstract}

We study a neutrino mass model with $A_4$ discrete flavor symmetry using a type-I seesaw mechanism. The inclusion of extra flavons in our model leads to the deviations from exact tribimaximal mixing pattern resulting in a nonzero $\theta_{13}$ consistent with the recent experimental results and a sum rule for light neutrino masses is also obtained. In this framework, a connection is established among the neutrino mixing angles- reactor mixing angle($\theta_{13}$), solar mixing angle($\theta_{12}$) and atmospheric mixing angle  ($\theta_{23}$). This model also allows us a prediction of Dirac CP-phase and Jarlskog parameter $J$. The octant of the atmospheric mixing angle $\theta_{23}$ occupies the lower octant. Our model prefers normal hierarchy (NH) than inverted hierarchy (IH). We use the parameter space of our model of neutrino masses to study the neutrinoless double beta decay parameter $m_{ee}$.

\end{abstract}             


\maketitle

%\tableofcontents

\section{\label{sec:1}Introduction}

The discovery of neutrino oscillations has triggered a lot of theoretical and experimental effort to understand the physics of lepton masses and mixing.
As flavor mixing happens due to the mismatch between the mass and flavor eigenstates, so neutrinos need to have small non-degenerate masses \cite{Maki:1962mu, Fusaoka:1972hn, Oberauer:1992vw}
Over the last twenty five years, numerous experiments on neutrino oscillation have taken place, resulting in the precise determination of oscillation parameters \cite{Forero:2012faj, Aartsen:2016psd,Gonzalez-Garcia:2012hef}. The discovery of neutrino oscillations in 1998 by Japanese Super-Kamiokande (SK) collaborators and Canadian Sudbury Neutrino Observatory collaborators was the first evidence of physics beyond the Standard Model. A few latest reviews on neutrino physics are placed in references \cite{Razzaghi:2023fsh,Frank:2022tbm, Ahn:2022ufs, Nath:2018ywc, Puyam:2022mej,Kang:2019gyu, Zhao:2017yvw, Buravov:2014dna, Nishi:2023ebi,  Hagedorn:2021ldq, Thapa:2023fxu,Bora:2023teg, Loualidi:2021qoq, Khatun:2020biw,Okada:2019lzv, Bhattacharyya:2022trp, Boruah:2021ktk, Girardi:2015slc,Mishra:2023cjc}.

The main factors influencing neutrino oscillation probabilities are the mass-squared differences and the mixing angles. Therefore, these parameters are determined in neutrino oscillation experiments. The experimental data has shown two large mixing angles, one is atmospheric mixing angle $\theta_{23}$ and another is solar mixing angle $\theta_{12}$ and one small mixing angle, called reactor mixing angle $\theta_{13}$. This pattern differs from quark mixing where all angles are small and the mixing matrix is close to identity.

The tribimaximal (TBM) mixing pattern is one of the most extensively used lepton mixing patterns obtained utilising discrete non-Abelian symmetries.

 \begin{equation}
    U_{TBM}=
    \begin{pmatrix}
    -\frac{\sqrt{2}}{\sqrt{3}} & \frac{1}{\sqrt{3}} & 0\\
    \frac{1}{\sqrt{6}} & \frac{1}{\sqrt{3}} & -\frac{1}{\sqrt{2}}\\
    \frac{1}{\sqrt{6}} & \frac{1}{\sqrt{3}} & \frac{1}{\sqrt{2}}
    \end{pmatrix}
\end{equation}

But TBM has been ruled out due to a non-zero reactor mixing angle, \cite{RENO:2012mkc, DoubleChooz:2011ymz}. One of the admired ways to achieve realistic mixing is through either its extensions or through modifications. In the concept of tribimaximal mixing (TBM), the angle $\theta_{13}$, which represents the degree of mixing in a reactor, is equal to zero and the CP phase $\delta_{CP}$, which characterizes the violation of symmetry between matter and antimatter, cannot be determined or has no specific value. Albeit in 2012 the Daya Bay Reactor Neutrino Experiment ($\sin^2 2\theta_{13} = 0.089 \pm 0.010 \pm 0.005 $) \cite{DayaBay:2012fng} and RENO Experiment $\sin^2 2\theta_{13}= 0.113 \pm 0.013 \pm 0.019 $ \cite{RENO:2012mkc} showed that $\theta_{13} \simeq 9 ^\circ$. Moreover, several neutrino oscillation experiments like MINOS \cite{MINOS:2011amj}, Double Chooz \cite{DoubleChooz:2011ymz}, T2K \cite{T2K:2011ypd}, measured consistent non-zero values for $\theta_{13}$. Other mixing angle values also show small deviations the TBM value.

The experiments investigating neutrino oscillation have discovered two mass-squared differences that vary significantly in their scales. The smaller mass squared differences, denoted $\Delta m_{21}^2= m^2_2 -m^2_1$, is positive and is of the order of $10^{-5} eV^2$ and the larger mass-squared difference,  $\Delta m_{31}^2= m^2_3 -m^2_1$, is of order $10^{-3} eV^2$ but its sign is unknown. It leads to two possible mass hierarchies for neutrinos: normal hierarchy (NH) in which $\Delta m_{31}^2$ is positive  and $m_1< m_2 <m_3$ and inverted hierarchy (IH) where  $m_3< m_1 <m_2$. Many experiments like INO \cite{blennow2012identifying, ghosh2013determining,ICAL:2015stm}, ICECube-PINGU \cite{Ribordy:2013xea, Quinn:2018zvt,Jia:2017oar,Winter:2013ema} and and long baseline
experiments \cite{agarwalla2014lbno, agarwalla2013neutrino} has the primary objective of determining the sign of $\Delta m_{31}^2$. The values of mixing angles and mass-squared differences and $\delta_{CP}$ from the global analysis of data is summarized in given in Table \ref{tab:1}.

\begin{table}[t]
\centering
  \begin{tabular}{ | l | c | r |}
    \hline
    Parameters & NH (3$\sigma$) & IH (3$\sigma$) \\ \hline
    $\Delta{m}^{2}_{21}[{10}^{-5}eV^{2}]$ & $6.82 \rightarrow 8.03$ & $6.82 \rightarrow 8.03$ \\ \hline
    $\Delta{m}^{2}_{31}[{10}^{-3}eV^{2}]$ & $2.428 \rightarrow 2.597$ & $-2.581 \rightarrow -2.408 $\\ \hline
    $\sin^{2}\theta_{12}$ & $0.270 \rightarrow 0.341$ & $0.270 \rightarrow 0.341$ \\ \hline
     $\sin^{2}\theta_{13}$ & $0.02029 \rightarrow 0.02391$ & $0.02047 \rightarrow 0.02396$ \\ \hline
    $\sin^{2}\theta_{23}$ & $0.405 \rightarrow 0.620$ & $0.410 \rightarrow 0.623$ \\ \hline
    $\delta_{CP}$ & $105 \rightarrow 405$ & $192 \rightarrow 361$ \\ \hline
  \end{tabular}
    \caption{The $3\sigma$ ranges of neutrino oscillation parameters from NuFIT 5.2 (2022) \cite{esteban2020fate}}
    \label{tab:1}
\end{table}

To elucidate the small masses of neutrinos in comparison to charged leptons and quarks, a new mechanism involving Majorana nature of neutrinos, called seesaw mechanism, was introduced in \cite{minkowski1977mu, yanagida1979proc, mohapatra1980neutrino, gell1979supergravity, achiman1978quark}. This method introduces right-handed neutrino companions with high-scale Majorana masses. In this mechanism, the right handed partners of neutrinos are introduced with Majorana masses at high scale. Furthermore, the neutrinos have Dirac masses of the order of charged lepton masses. Also, there are some other frameworks beyond the standard model (BSM) that can explain the origin of neutrino masses, for examples, Supersymmetry \cite{ma1999supersymmetry},
Minimal Supersymmetric Standard Model (MSSM) \cite{Csaki:1996ks}, Minimal seesaw model\cite{kang2021low}, Inverse seesaw model\cite{Hue:2021zyw}, Next-to-Minimal Supersymmetric Standard Model (NMSSM) \cite{Ellwanger:2009dp}, String theory \cite{ibanez2012string}, models based on extra dimensions \cite{Arkani-Hamed:1998wuz}, Radiative Seesaw Mechanism \cite{Thapa:2022fhv,Ma:2006km} and also some other models. And also various models based on non-abelian discrete flavor symmetries \cite{King:2013eh} like $A_4$ \cite{Ishimori:2010au, Nguyen:2020ehj, Ganguly:2022qxj, Behera:2020sfe, Vien:2022cxr, Ma:2005qf, Ma:2015fpa}, $S_3$ \cite{ma2004non}, $S_4$ \cite{Bazzocchi:2012st, Vien:2020aya, Thapa:2021ehj, Ma:2005mw, dev2015non, Grossman:2014oqa}, $\Delta_{27}$ \cite{Ma:2007wu, hernandez20163, de2007neutrino, CarcamoHernandez:2020udg}, $\Delta_{54}$ \cite{Bora:2023teg, ishimori2009lepton, carballo2016delta} etc. have been
proposed to obtain tribimaximal mixing (TBM) and deviation from TBM.

The mixing between the neutrino flavour eigenstates and their mass eigenstates is encoded by the commonly used PMNS matrix. This PMNS matrix is parameterized in a three-flavoured paradigm using three mixing angles and three CP phases as given below:

\begin{equation}
\label{eq:1}
    U_{PMNS}=
    \begin{pmatrix}
    c_{12} c_{13} & s_{12} c_{13} & s_{13}e^{-i \delta}\\
    -s_{12} c_{23}- c_{12} s_{23} s_{13} e^{ i \delta} & c_{12} c_{23} - s_{12} s_{23} s_{13} e^{i \delta} & s_{23} c_{13}\\
    s_{12} s_{23} - c_{12} c_{23} s_{13} e^{ i \delta} & -c_{12} s_{23} -s_{12} c_{23} s_{13} e^{ i \delta} & \ c_{23} c_{13}
    \end{pmatrix}
    \cdot U_{Maj}
\end{equation}
where, $ c_{ij}=\cos{\theta_{ij}}$, $s_{ij}=\sin{\theta_{ij}}$. The diagonal matrix $U_{Maj}= diag (1, e^{ i \alpha}, e^{ i (\beta+\gamma)})$ contains the Majorana CP phases, $\alpha$,  $\beta$ which become observable in case the neutrinos behave as Majorana particles. Identifying neutrinoless double beta decay will probably be necessary to prove that neutrinos are Majorana particles. Such decays have not yet been seen. Here, symmetry will play an important role in explaining these problems.
In order to account for the fact that neutrino mass is zero within the standard model (SM) \cite{Liu:2016oph}, it becomes necessary to develop a new framework that goes beyond the standard model. This entails incorporating a new symmetry and creating a mechanism that generates non-zero masses for neutrinos.

In this study, we put forward a model for neutrino masses to provide an explanation for the observed non-zero value of $\theta_{13}$, as well as the existing data on neutrino masses and mixings. To get the deviation from exact TBM neutrino mixing pattern, we have extended the flavon sector of Altarelli-Feruglio (A-F) \cite{Altarelli:2005yx, Altarelli:2010gt} model by introducing extra flavons $\xi^\prime$, $\xi^{\prime\prime}$ and $\rho$ which transform as $1^\prime$, $1^{\prime\prime}$ and 1 respectively under $A_4$.
Here, type-I see-saw framework \cite{Borah:2017dmk, Crivellin:2022cve} is utilized to construct the model. Also, we incorporated a $Z_2\times Z_3$ symmetry as well, which serves the purpose of avoiding undesired terms.

The content material of our paper is organised as follows: In section 2, we give the overview of the framework of our model by specifying the fields involved and their transformation properties under the symmetries imposed. In section 3, we do numerical analysis and study the results for the neutrino phenomenology. We finally conclude our work in section 4.

\section{\label{sec:f}Framework of the Model}

Here we provide a concise overview of the ways in which the non-Abelian discrete symmetry $A_4$ group can be represented \cite{Altarelli:2010gt,Kobayashi:2022moq}. $A_4$ is a group of even permutations of four objects and it has 12 elements (12= $\frac{4!}{2}$). The group known as the tetrahedron group, or sometimes referred to as the group of orientation-preserving symmetries of a regular tetrahedron, is characterized by its ability to describe the symmetries of this particular geometric shape. This can be generated by two basic permutations S and T having properties $S^2=T^3=(ST)^3=1$. This group representations of $A_4$ include three one-dimensional unitary representations $1$, $1^\prime$,  $1^{\prime\prime}$ with the generators S and T given, respectively as follows:
$$1: S=1, T=1$$
$$1^\prime: S=1, T=\omega^2$$
$$1^{\prime\prime}: S=1, T=\omega$$
and a three dimensional unitary representation with the generators\footnote {Here the generator T has been chosen to be diagonal}
\begin{equation}
    T=
    \begin{pmatrix}
    1 & 0 & 0\\
    0 & \omega^2 & 0\\
    0 & 0 & \omega
    \end{pmatrix}
\end{equation}

\begin{equation}
    S=\frac{1}{3}
    \begin{pmatrix}
    -1 & 2 & 2\\
    2 & -1 & 2\\
    2 & 2 & -1
    \end{pmatrix}
\end{equation}\\.
Here $\omega$ is the cubic root of unity, $\omega=exp(i2\pi)$, so that $1+\omega+\omega^2 =0$.\\
The multiplication rules corresponding to the specific basis of two generators S and T are as follows:
$$1\times1=1$$
$$1^{\prime\prime}\times1^\prime =1$$
$$1^\prime\times\ 1^{\prime\prime}=1$$
$$3\times3=3+ 3_A + 1+ 1^\prime +1^{\prime\prime}$$

For two triplets\\
$$a= (a_1,a_2, a_3)$$
$$b= (b_1,b_2, b_3)$$
we can write
$$1 \equiv (ab) = a_1 b_1 +a_2 b_3 +a_3 b_2$$
$$1^\prime \equiv (ab)^\prime=a_3 b_3 +a_1 b_2 +a_2 b_1$$
$$1^{\prime\prime} \equiv (ab)^{\prime\prime}=a_2 b_2 +a_1 b_3 +a_3 b_1$$

Here 1 is symmetric under the exchange of second and third elements of a and b, $1^\prime$ is symmetric under the exchange of the first and second elements while $1^{\prime\prime}$ is symmetric under the exchange of first and third elements.
$$3 \equiv (ab)_S=\frac{1}{3}(2a_1 b_1 -a_2 b_3 -a_3 b_2, 2a_3 b_3 -a_1 b_2 -a_2 b_1, 2a_2 b_2 -a_1 b_3 -a_3 b_1)$$
$$3_A \equiv (ab)_A =\frac{1}{3}(a_2 b_3 -a_3 b_2, a_1 b_2 -a_2 b_1, a_1 b_3 -a_3 b_1)$$

Here 3 is symmetric and $3_A$ is anti-symmetric. For the symmetric case, we notice that the first element has 2-3 exchange symmetry, the second element has 1-2 exchange symmetry and the third element has 1-3 exchange symmetry.

\begin{table}[t]
    \centering
    \begin{tabular}{c c c c c c c c c c c c c c}
    \hline
       \textrm{Field}  &  l & $e^c$ & $\mu^c$ & $\tau^c$ & $h_u$ & $h_d$ & $\nu^c$ & $\Phi_S$ & $\Phi_T$ & $\xi$ & $\xi^\prime$ & $\xi^{\prime\prime}$ & $\rho$ \\
     \hline
     
     \textrm{SU(2)}  &  2 & 1 & 1 & 1 & 2 & 2 & 1 & 1 & 1 & 1 & 1 & 1 & 1\\
     
     \textrm {A}$_4$  &  3 & 1 & $1^{\prime\prime}$ & $1^\prime$ & 1 & 1 & 3 & 3 & 3 & 1 & $1^{\prime\prime}$ & $1^\prime$ & 1 \\
     
    \textrm{Z}$_2$  &  1 & -1 & -1 & -1 & 1 & 1 & 1 & 1 & -1 & 1 & 1 & 1 & 1 \\
    
    \textrm{Z}$_3$  &  $\omega^2$ & $\omega$ & $\omega$ & $\omega$ & 1 & 1 & 1 & $\omega$ & 1 & $\omega$ & $\omega$ & $\omega$ & $\omega$ \\
    \hline
    \end{tabular}
    \caption{Full particle content of our model}
    \label{tab:2}
\end{table}

The particle content and their charge assignment under the symmetry group is given in Table \ref{tab:2}. The left-handed lepton doublets l and right-handed charged leptons ($e^c, \mu^c, \tau^c$) are assigned to triplet and singlet ($1,  1^{\prime \prime}, 1^\prime $ ) representation under A$_4$ respectively and other particles transform as shown in Table-II. Here, $h_u$ and $h_d$ are the standard Higgs doublets which remain invariant under $A_4$. The right-handed neutrino field $\nu^c$ is assigned to the triplet representation under $A_4$ flavor symmetry. There are six $SU(2)\otimes U_Y (1)$ Higgs singlets, four ($\xi$, $\xi^\prime$, $\xi^{\prime\prime}$ and $\rho$) of which singlets under $A_4$ and two ($\Phi_T$ and $\Phi_S$) of which transform as triplets.

Consequently, the invariant Yukawa Lagrangian is as follows:

\begin{multline}
 \mathcal{-L}= \frac{y_e}{\Lambda}(l \Phi_T)_1 h_d e^c+
   \frac{y_\mu}{\Lambda}(l \Phi_T)_{1^\prime} h_d \mu^c+
    \frac{y_\tau}{\Lambda}(l \Phi_T)_{1^{\prime\prime}} h_d \tau^c+ 
     \frac{y_1}{\Lambda}\xi_1 (l h_u \nu^c)_1+
     \frac{y_2}{\Lambda} {\xi_2}_{1^{\prime \prime}} (l h_u \nu^c)_{1^\prime}+ \\
      \frac{y_3}{\Lambda} {\xi_3}_{1^\prime} (l h_u \nu^c)_{1^{\prime\prime}} 
     + \frac{y_4}{\Lambda} {\rho}_1 (l h_u \nu^c)_1+
     \frac{y_a}{\Lambda}\Phi_S (l h_u \nu^c)_A+
     \frac{y_b}{\Lambda}\Phi_S (l h_u \nu^c)_S+ \frac{1}{2} M_N (\nu^c \nu^c) +h.c.
\end{multline}

The terms $y_e$, $y_\mu$, $y_\tau$, $y_1$, $y_2$, $y_3$, $y_4$, $y_a$ and $y_b$ are coupling constant and $\Lambda$ is the cut-off scale of the theorys. We assume $\Phi_T$ does not couple to the Majorana mass matrix and $\Phi_S$ does not couple to the charged leptons. After spontaneous symmetry breaking of flavour and electroweak symmetry we obtain the mass matrices for the charged leptons and neutrinos. We assume the vacuum alignment of $\langle \Phi_T \rangle=(v_T ,0,0)$ and $\langle \Phi_S \rangle=(v_s ,v_s,v_s)$. Also, $v_u, v_d$ are the VEVs of $\langle h_u \rangle $, $\langle h_d \rangle$ and $u_1, u_2, u_3, u_4$ are the VEVs of  $\langle \xi_1 \rangle $, $\langle \xi_2 \rangle $, $\langle \xi_3 \rangle $,$\langle \rho \rangle $ respectively.

The VEV pattern of the $A_4$ triplets, which is considered in our model, has been thoroughly examined in numerous $A_4$ models like \cite{Barman:2022hyq,Altarelli:2010gt}.

The charged lepton mass matrix is given as 
\begin{equation}
    M_l= \frac{v_d v_T}{\Lambda}
    \begin{pmatrix}
    y_e & 0 & 0\\
    0 & y_\mu & 0\\
    0 & 0 & y_\tau
    \end{pmatrix}
\end{equation}

where, $v_d$ and $v_T$ are the VEVs of $h_d$ and $\Phi_T$ respectively. 

The structure of the Majorana neutrino mass matrix:
\begin{equation}
    M_R= 
\begin{pmatrix}
    M_N & 0 & 0\\
    0 & 0 & M_N\\
    0 & M_N & 0
    \end{pmatrix}
\end{equation}

The form of the Dirac mass matrix:

\begin{equation}
    M_D= 
\begin{pmatrix}
    \frac{2b}{3}+c+f & -\frac{a}{3}-\frac{b}{3}+d & -\frac{a}{3}-\frac{b}{3}+e\\
    \frac{a}{3}-\frac{b}{3}+d & \frac{2b}{3}+e & -\frac{a}{3}-\frac{b}{3}+c+f\\
   \frac{a}{3}-\frac{b}{3}+e & \frac{a}{3}-\frac{b}{3}+c+f & \frac{2b}{3}+d
    \end{pmatrix}
\end{equation}
Where, $a=\frac{y_a v_u v_s}{\Lambda}$, $b=\frac{y_b v_u v_s}{\Lambda}$, $c=\frac{y_1 v_u u_1}{\Lambda}$, $d=\frac{y_2 v_u u_2}{\Lambda}$, $e=\frac{y_3 v_u u_3}{\Lambda}$ and $f=\frac{y_4 v_u u_4}{\Lambda}$.

The Type-I seesaw method is used to determine the effective neutrino mass matrix
$m_\nu= M^T_D M^{-1}_R M_D $
\begin{equation}
    m_\nu= 
\begin{pmatrix}
    m_{11} & m_{12} & m_{13}\\
    m_{12} & m_{22} & m_{23}\\
    m_{13} & m_{23} & m_{33}
    \end{pmatrix}
\end{equation}

Where, 

$m_{11}=\frac{1}{M_N}[2(\frac{a}{3}-\frac{b}{3}+d)(\frac{a}{3}-\frac{b}{3}+e)+(\frac{2b}{3}+c+f)^2]$

$m_{12}=m_{21}=\frac{1}{M_N}[(\frac{a}{3}-\frac{b}{3}+e)(\frac{2b}{3}+e)+(\frac{a}{3}-\frac{b}{3}+d)(\frac{a}{3}-\frac{b}{3}+c+f)+(-\frac{a}{3}-\frac{b}{3}+d)(\frac{2b}{3}+c+f)]$

$m_{13}=m_{31}=\frac{1}{M_N}[(\frac{a}{3}-\frac{b}{3}+d)(\frac{2b}{3}+d)+(\frac{a}{3}-\frac{b}{3}+e)(-\frac{a}{3}-\frac{b}{3}+c+f)+(-\frac{a}{3}-\frac{b}{3}+e)(\frac{2b}{3}+c+f)]$

$m_{22}=\frac{1}{M_N}[(-\frac{a}{3}-\frac{b}{3}+d)^2+2(\frac{2b}{3}+e)(\frac{a}{3}-\frac{b}{3}+c+f)]$

$m_{23}=m_{32}=\frac{1}{M_N}[(-\frac{a}{3}-\frac{b}{3}+d)(-\frac{a}{3}-\frac{b}{3}+e)+(\frac{2b}{3}+d)(\frac{2b}{3}+e)+(-\frac{a}{3}-\frac{b}{3}+c+f)(\frac{a}{3}-\frac{b}{3}+c+f)]$

$m_{33}=\frac{1}{M_N}[(-\frac{a}{3}-\frac{b}{3}+e)^2+2(\frac{2b}{3}+d)(-\frac{a}{3}-\frac{b}{3}+c+f)]$

To explain the smallness of active neutrino masses, we consider the heavy neutrino with masses $M_N \approx 10^{14}$ GeV. We can assume $c \simeq d \simeq e \simeq f$. This is a reasonable assumption to make since the phenomenology does not change drastically unless the VEVs of the singlet Higgs vary by a huge amount \cite{Barman:2022hyq,Brahmachari:2008fn}. Thus, the neutrino mass matrix becomes with new matrix elements:

$m^\prime_{11}=\frac{1}{9M_N}[2(a^2-2ab+3b^2+6(a+b)c+ 27c^2)]$

$m^\prime_{12}=m^\prime_{21}=\frac{1}{9M_N}[a^2-2a(b-3c)-3(b-3c)(b+5c)]$

$m^\prime_{13}=m^\prime_{31}=\frac{1}{M_N}[-a^2+3(b-3c)(b+5c)]$

$m^\prime_{22}=\frac{1}{9M_N}[a^2+6ab-3b^2+12bc+45c^2]$

$m^\prime_{23}=m^\prime_{32}=\frac{1}{9M_N}[2b(a+3b)-6(a+b)c+54c^2]$

$m^\prime_{33}=\frac{1}{9M_N}[a^2-3b^2+12bc+45c^2-2a(b+6c)]$

In section 3, we give the detailed phenomenological analysis  on various neutrino oscillation parameters. Further, we present a numerical study of neutrinoless double-beta decay considering the allowed parameter space of the model.

\section{Numerical Analysis and results}

\noindent The neutrino mass matrix $m_\nu$ can be diagonalized by the PMNS matrix $U$ as
\begin{equation}
    \label{eq:12}
    U^\dagger m_\nu U^* = \textrm{diag(}m_1, m_2, m_3 \textrm{)}
\end{equation}.

We can numerically calculate $U$ using the relation $U^\dagger h U = \textrm{diag(}m_1^2, m_2^2, m_3^2 \textrm{)}$, where, $h=m_\nu m^\dagger_\nu$. The neutrino oscillation parameters $\theta_{12}$, $\theta_{13}$, $\theta_{23}$ and $\delta_{CP}$ can be obtained from $U$ as
\begin{equation}
    \label{eq:13}
    s_{12}^2 = \frac{\lvert U_{12}\rvert ^2}{1 - \lvert U_{13}\rvert ^2}, ~~~~~~ s_{13}^2 = \lvert U_{13}\rvert ^2, ~~~~~~ s_{23}^2 = \frac{\lvert U_{23}\rvert ^2}{1 - \lvert U_{13}\rvert ^2},
\end{equation}

and $\delta$ may be given by
\begin{equation}
    \label{eq:14}
    \delta = \textrm{sin}^{-1}\left(\frac{8 \, \textrm{Im(}h_{12}h_{23}h_{31}\textrm{)}}{P}\right)
\end{equation}
with 
\begin{equation}
    \label{eq:15}
     P = (m_2^2-m_1^2)(m_3^2-m_2^2)(m_3^2-m_1^2)\sin 2\theta_{12} \sin 2\theta_{23} \sin 2\theta_{13} \cos \theta_{13}
\end{equation}

For the comparison of theoretical neutrino mixing parameters with the latest experimental data \cite{esteban2020fate}, the $A_4$ model is fitted to the experimental data by minimizing the following $\chi^2$ function:

\begin{equation}
	\label{eq:16}
	\chi^2 = \sum_{i}\left(\frac{\lambda_i^{model} - \lambda_i^{expt}}{\Delta \lambda_i}\right)^2.
\end{equation}
where $\lambda_i^{model}$ is the $i^{th}$ observable predicted by the model, $\lambda_i^{expt}$ stands for the $i^{th}$ experimental best-fit value and $\Delta \lambda_i$ is the $1\sigma$ range of the $i^{th}$ observable.

\begin{figure}[t]
     \centering
     \begin{subfigure}[b]{0.46\textwidth}
         \centering
         \includegraphics[width=\textwidth]{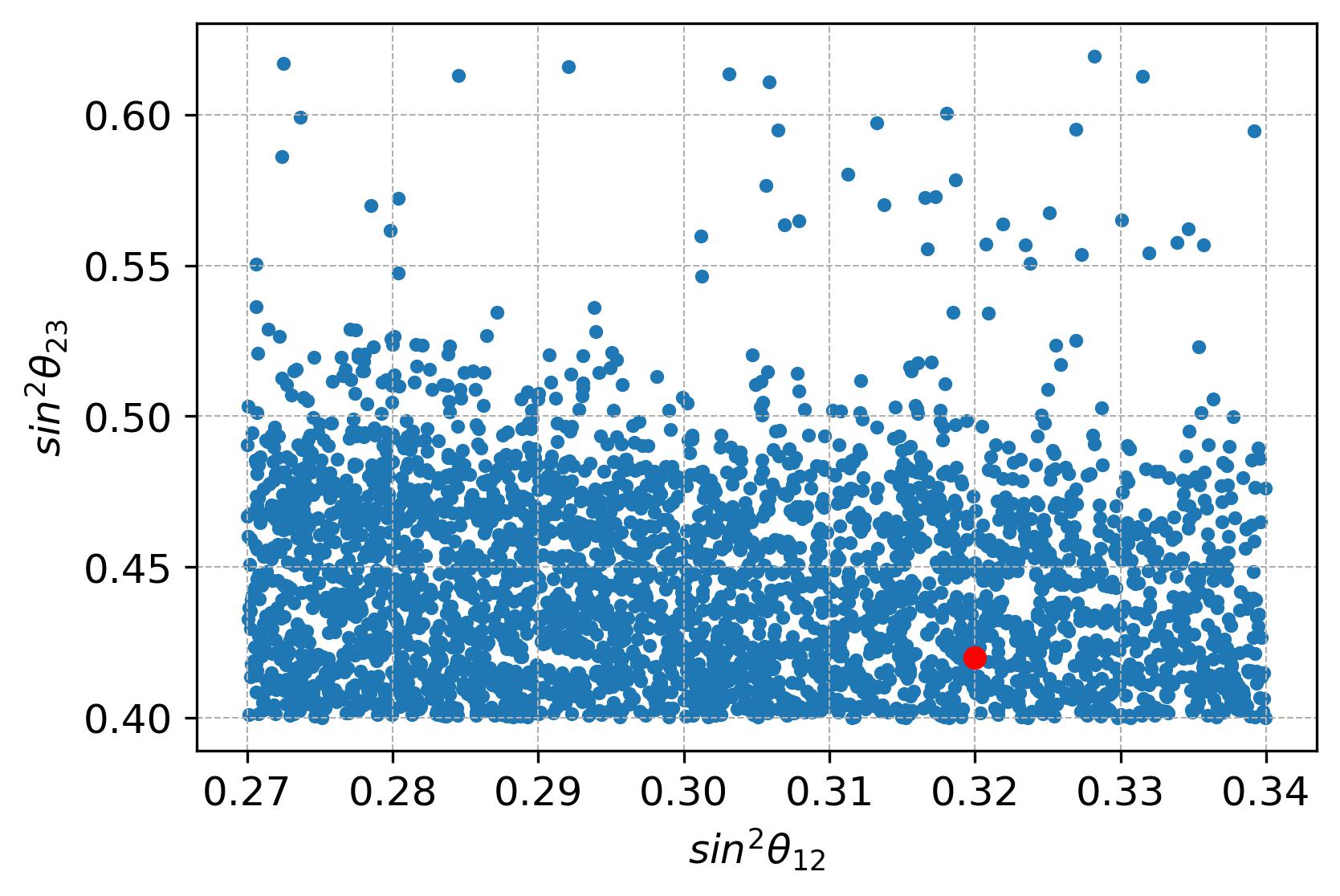}
     \end{subfigure}
     \hfill
     \begin{subfigure}[b]{0.46\textwidth}
         \centering
         \includegraphics[width=\textwidth]{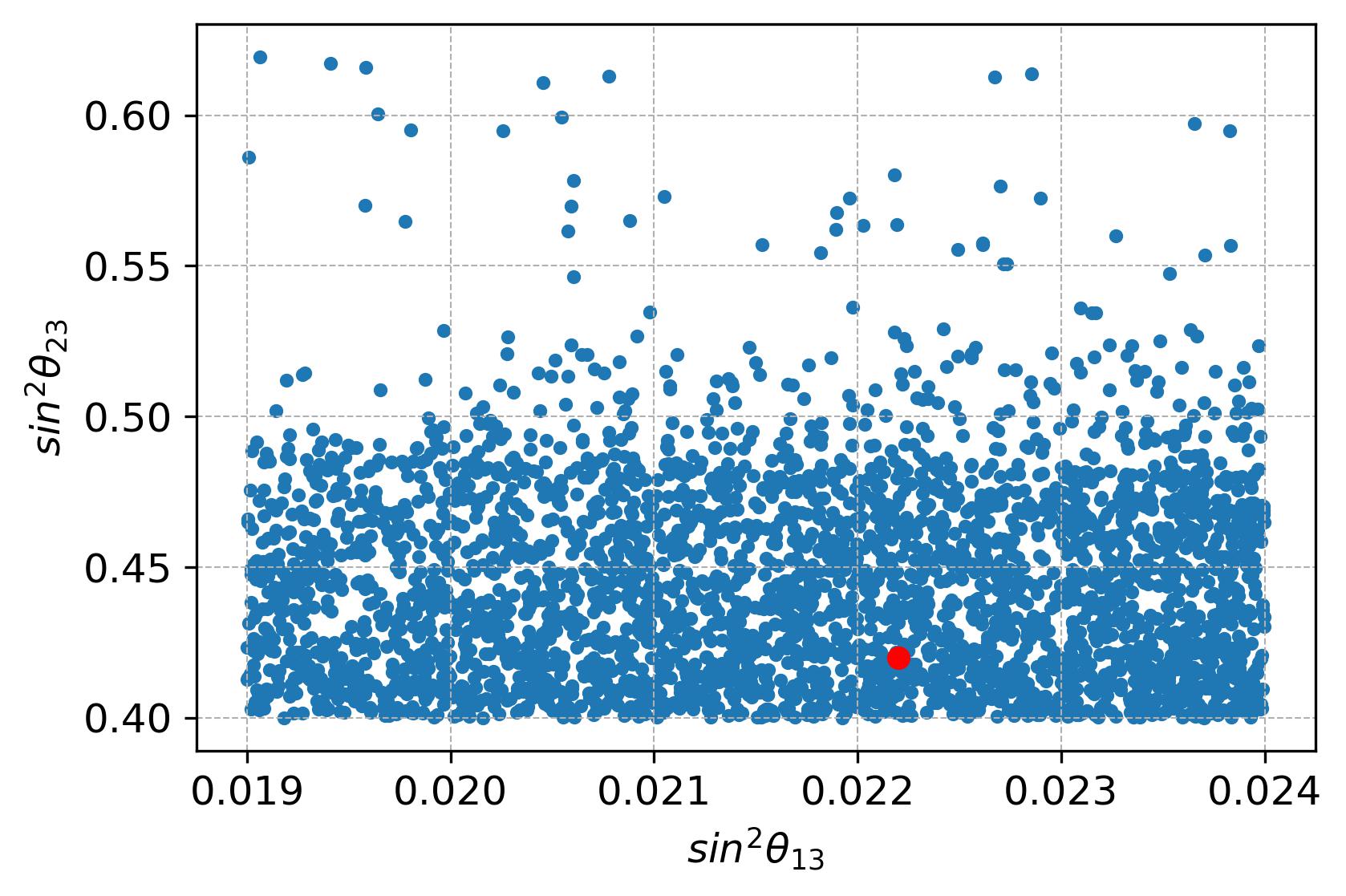}
     \end{subfigure}

     \vspace{1em}
     \begin{subfigure}[b]{0.46\textwidth}
         \centering
         \includegraphics[width=\textwidth]{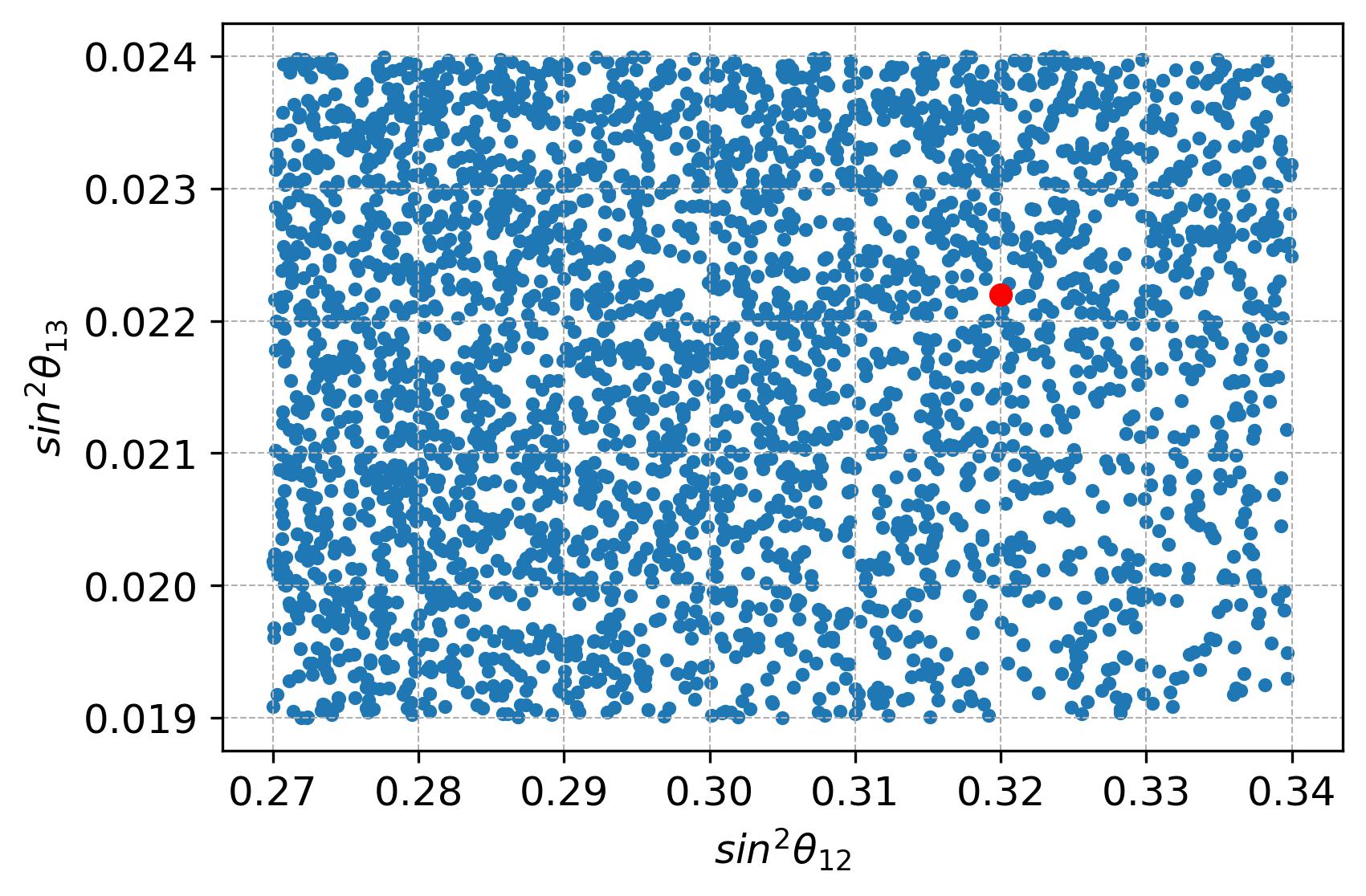}
     \end{subfigure}
     \hfill
     \begin{subfigure}[b]{0.46\textwidth}
         \centering
         \includegraphics[width=\textwidth]{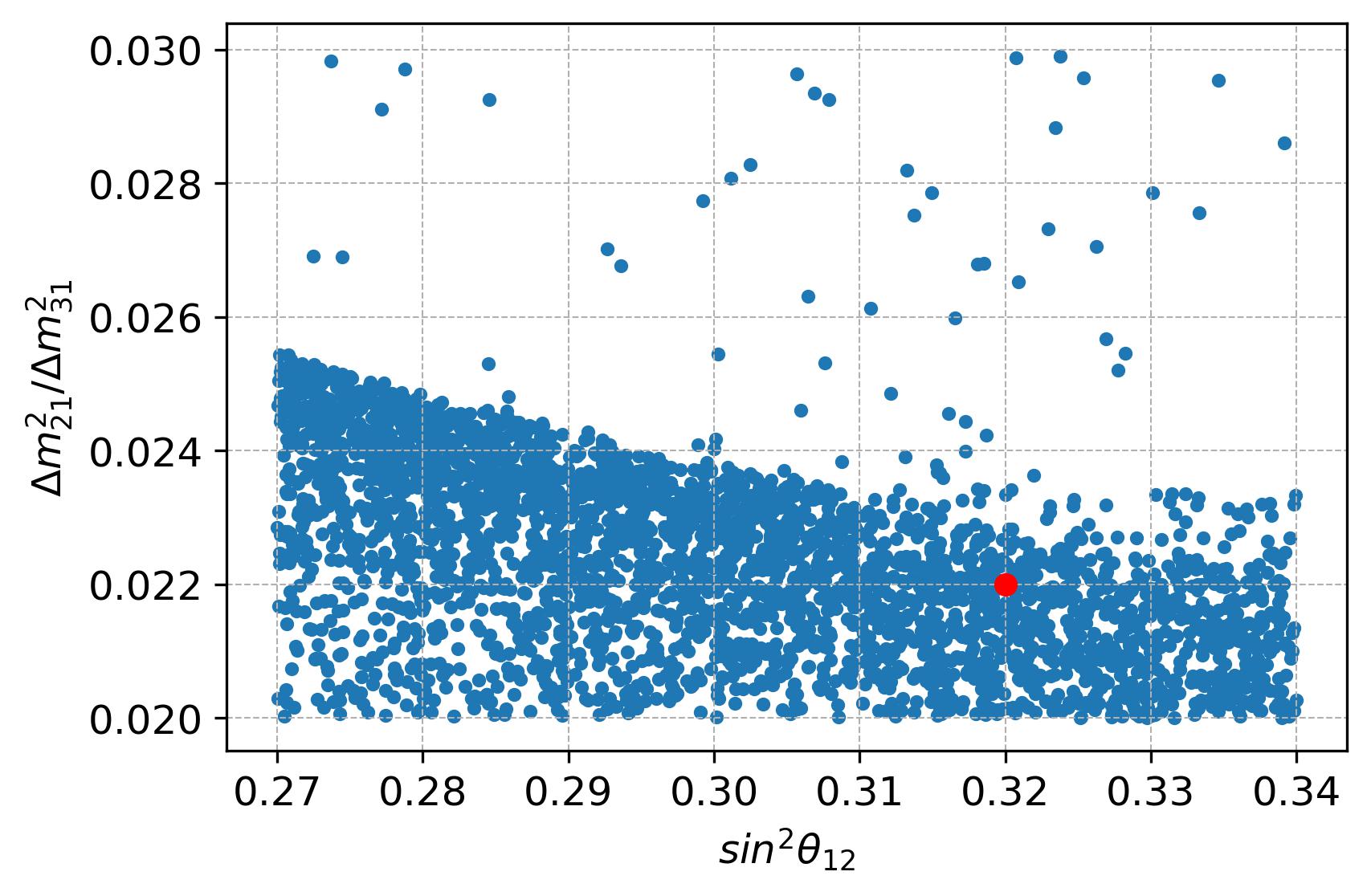}
     \end{subfigure}
    \caption{Correlation among the neutrino oscillation parameters $\sin^2\theta_{12}$, $\sin^2\theta_{23}$,
    $\frac{\Delta m^2_{21}}{\Delta m^2_{31}}$ and $\sin^2\theta_{13}$
    }
    \label{fig:1}
\end{figure}

\begin{figure}[!ht]
     \centering
     \begin{subfigure}[b]{0.46\textwidth}
         \centering
         \includegraphics[width=\textwidth]{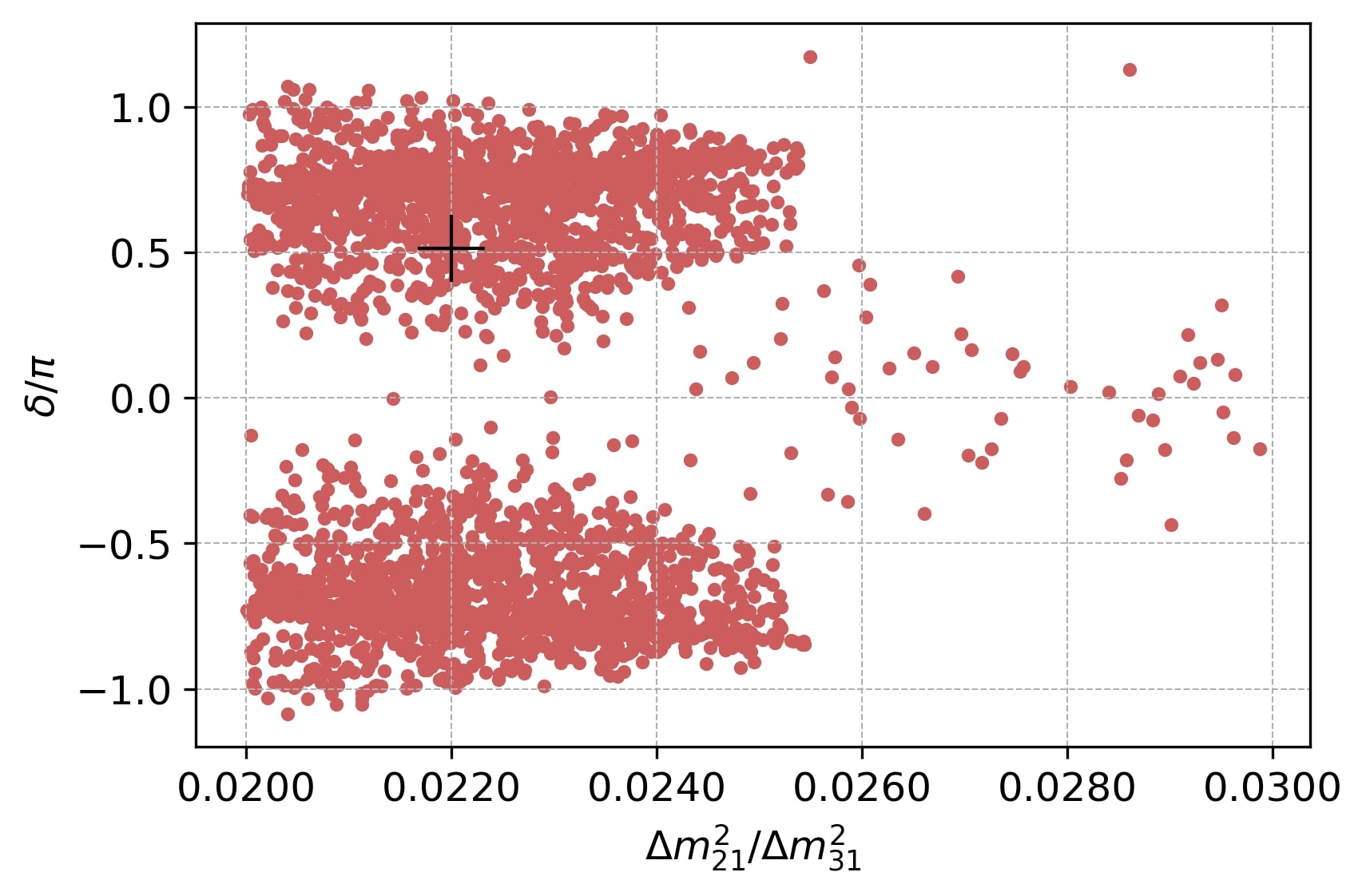}
     \end{subfigure}
     \hfill
     \begin{subfigure}[b]{0.46\textwidth}
         \centering
         \includegraphics[width=\textwidth]{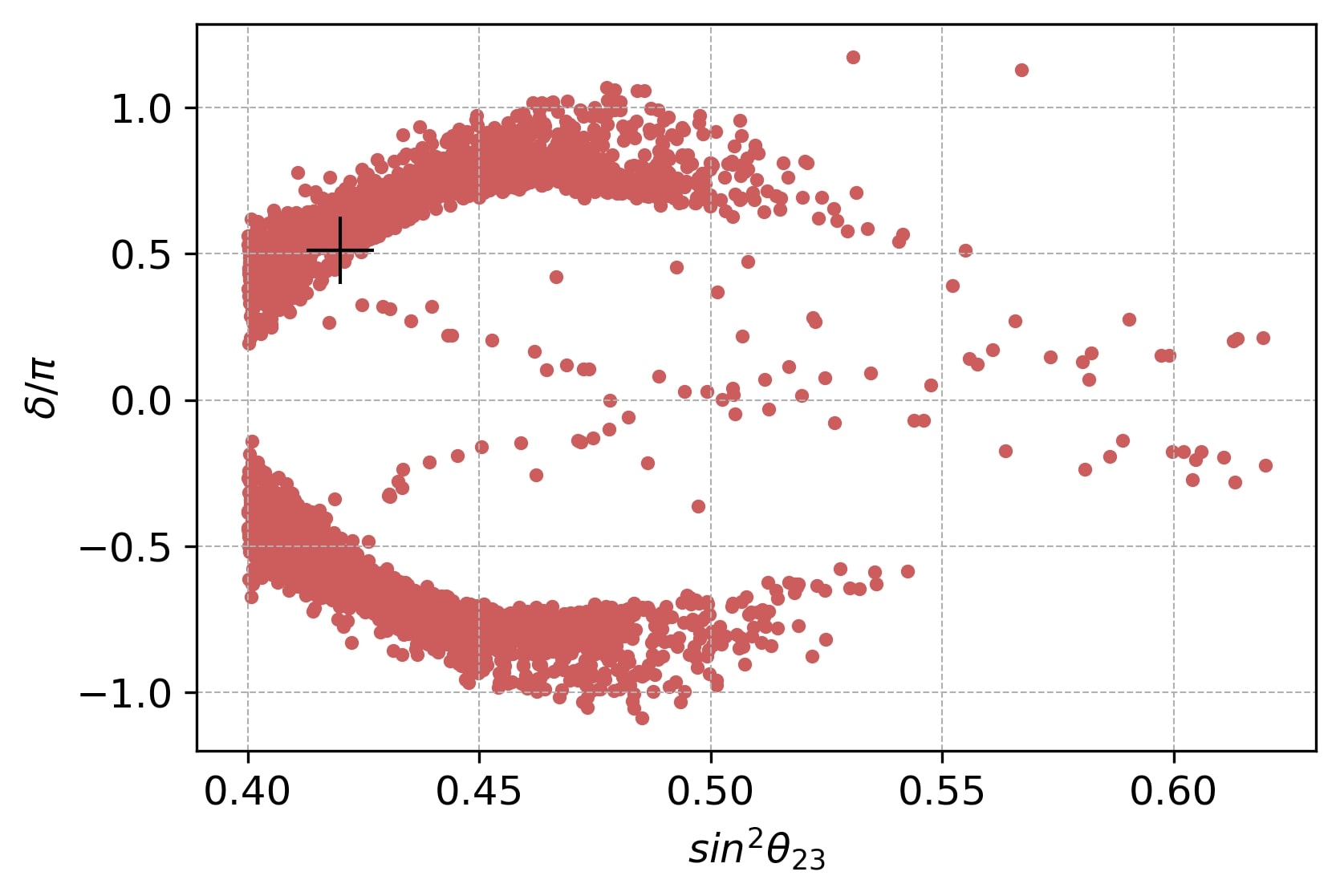}
     \end{subfigure}

     \vspace{1em}
     \begin{subfigure}[b]{0.46\textwidth}
         \centering
         \includegraphics[width=\textwidth]{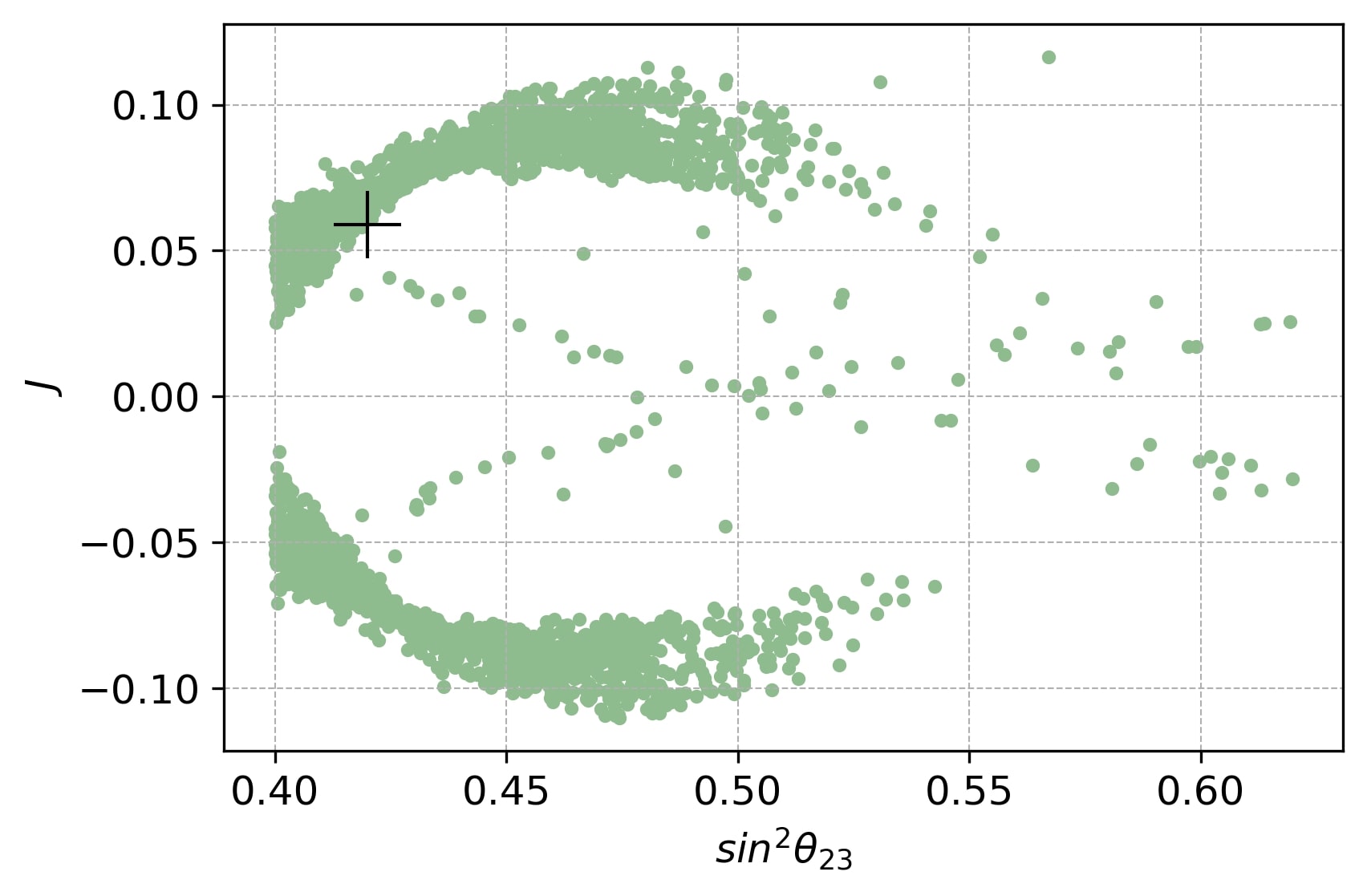}
     \end{subfigure}
     \hfill
     \begin{subfigure}[b]{0.46\textwidth}
         \centering
         \includegraphics[width=\textwidth]{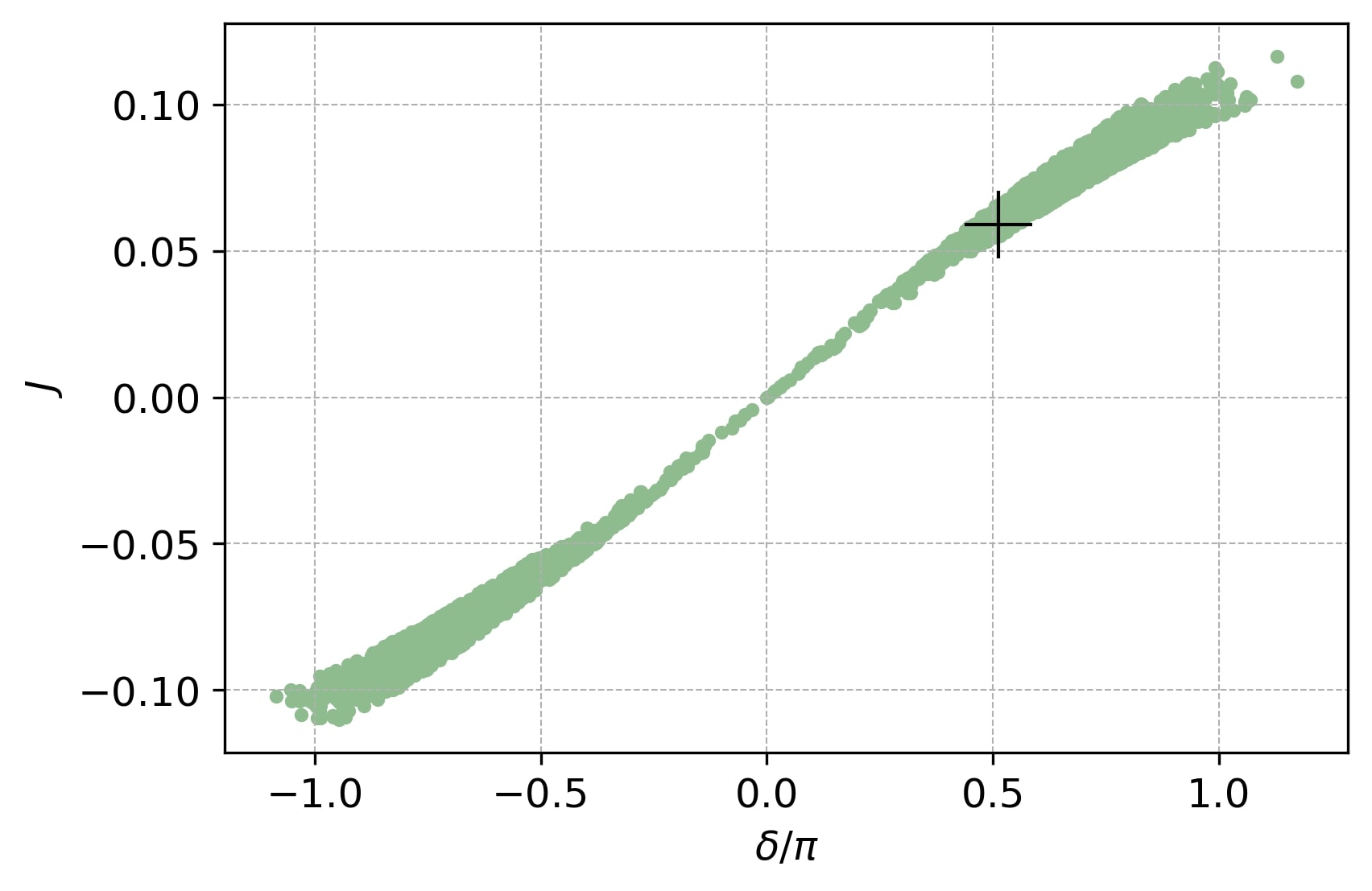}
     \end{subfigure}
    \caption{Correlation among the mixing angles, ratio of mass-squared differences  $\frac{\Delta m^2_{21}}{\Delta m^2_{31}}$, Dirac CP phase and  Jarlskog parameter J. }
    \label{fig:2}
\end{figure}

In Fig. \ref{fig:1}, correlation among the neutrino oscillation parameters $\sin^2\theta_{12}$, $\sin^2\theta_{23}$,  $\frac{\Delta m^2_{21}}{\Delta m^2_{31}}$ and $\sin^2\theta_{13}$ for NH has shown, which is constrained using the 3$\sigma$ bound on neutrino oscillation data. We can see that there is a high correlation among different parameters of the model. Fig. \ref{fig:2} shows the correlation among the mixing angles, ratio of mass-squared differences, Jarlskog parameter $J$ and Dirac CP phase for NH.

The calculated best fit values of $\sin^2 \theta_{12}$, $\sin^2 \theta_{13}$ and $\sin^2 \theta_{23}$ are (0.342, 0.0238, 0.556) which are within the 3 $\sigma$ range of experimental values. Other parameters such as $\Delta m_{21}^2$, $\Delta m_{31}^2$ and $\delta_{CP}$ have their best-fit values, corresponding to $\chi^2$-minimum, at ($7.425 \times 10^{-5} eV^2$, $2.56 \times 10^{-3} eV^2$, $-0.358 \pi$) respectively, which perfectly agreed with the latest observed neutrino oscillation experimental data. Thus, the model defined here, clearly shows the deviation from exact tri-bimaximal mixing.

\begin{figure}[t]
     \centering
     \begin{subfigure}[b]{0.46\textwidth}
         \centering
         \includegraphics[width=\textwidth]{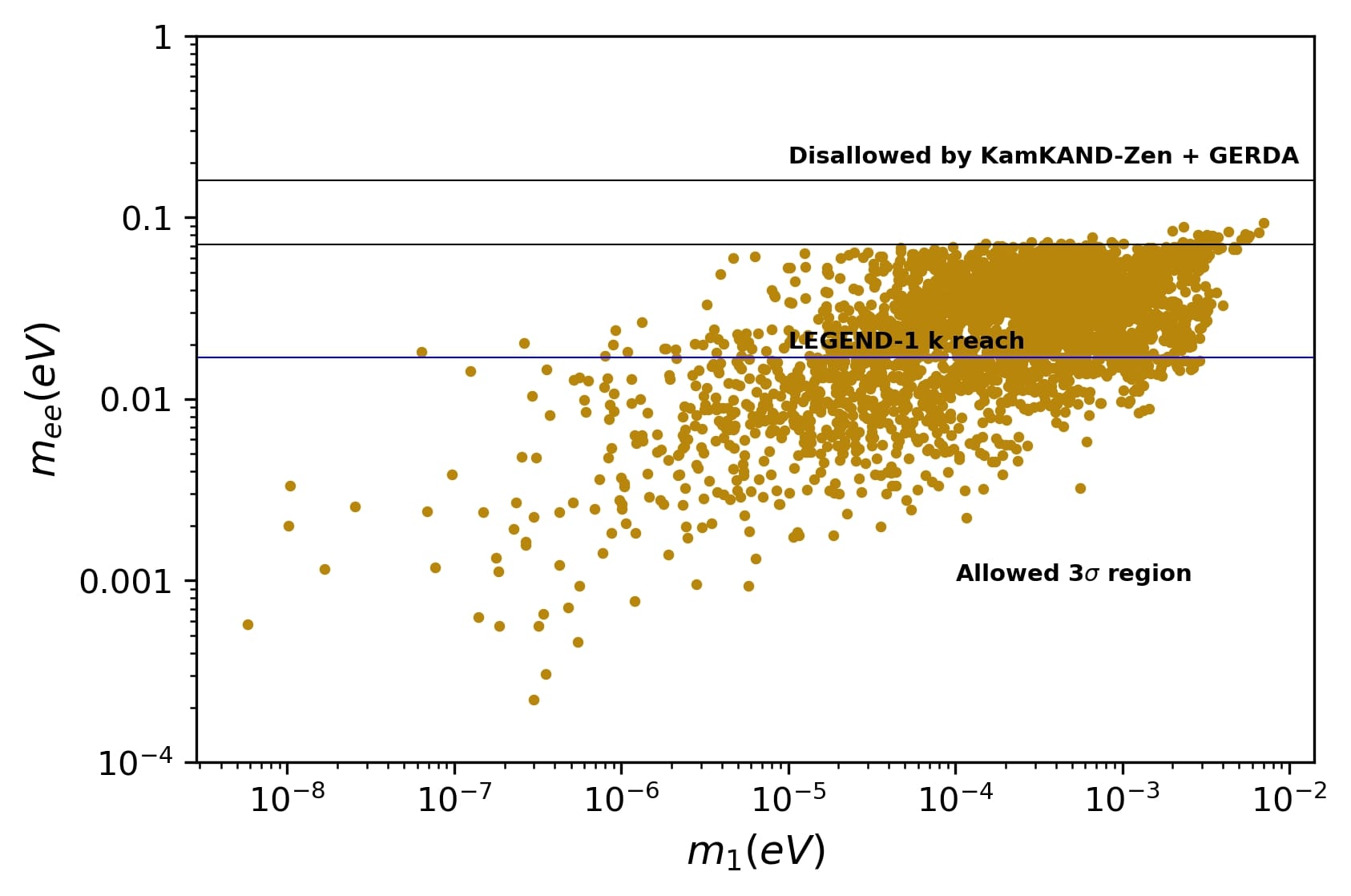}
     \end{subfigure}
     \hfill
    
    \caption{Variation of effective Majorana neutrino mass with lighest neutrino mass in NH with the KamLAND-Zen-Gerda bound on the effective mass.}
    \label{fig:3}
\end{figure}

\noindent \textbf {Neutrinoless double beta decay (NDBD):}

Up until now, the question of whether neutrinos belong to the Dirac or Majorana category remains unanswered. If they are of the Majorana type, the investigation of Neutrinoless Double Beta Decay (NDBD) becomes highly significant. Several ongoing experiments are being conducted to ascertain the Majorana nature of neutrinos. The effective mass that controls this process is furnished by

\begin{equation}
m_{\beta\beta}= U^2_{Li} m_i
\end{equation}
where $U_{Li}$ are the elements of the first row of the neutrino mixing matrix $U_{PMNS}$
(Eq.\ref{eq:1}) which is dependent on known parameters $\theta_{12}$, $\theta_{13}$ and the unknown Majorana phases $\alpha$ and $\beta$. $U_{PMNS}$ is the diagonalizing matrix of the light neutrino mass matrix $m_\nu$ so that,
\begin{equation}
m_\nu= U_{PMNS} M^{(diag)}_\nu U^T_{PMNS}
\end{equation}
where, $m^{(diag)}_\nu$ =diag($m_1$, $m_2$, $m_3$). The effective Majorana mass can be parameterized using the diagonalizing matrix elements and the mass eigen values as follows:
\begin{equation}
m_{\beta\beta}= m_1 c_{12}^2 c^2_{13}+ m_2 s^2_{12} c^2_{13} e^{2 i \alpha} + m_3 s^2_{13} e^{2i \beta}
\end{equation}

Using the constrained parameter space we have evaluated the value of $m_{\beta\beta}$ for our model. The variation of $m_{\beta\beta}$
with lightest neutrino mass is shown in figure \ref{fig:3}. The sensitivity reach of NDBD experiments like KamLAND-Zen \cite{Gando:2020cxo, Biller:2021bqx}, GERDA \cite{DiMarco:2020mdk, GERDA:2017wlm, DAndrea:2020ezp},
LEGEND-1k \cite{abgrall2021legend} is also shown in figure \ref{fig:3}. 
$m_{\beta\beta}$ is found to be well within the sensitivity reach of these NDBD experiments for our model.

\section{Conclusions}

We have developed a flavon-symmetric $A_4 \times Z_2 \times Z_3$ model using seesaw type-I mechanism. This model aims to explain the recent experimental data on neutrino oscillation, which deviate from the Tribimaximal neutrino mixing pattern. The inclusion of the cyclic $Z_2\times Z_3$ symmetric component has been done to remove undesired terms during the computations. The computed values unequivocally demonstrated that the neutrino mixing parameters deviate from the exact Tri-bimaximal neutrino mixing matrix. The resulting mass matrices give predictions for the neutrino oscillation parameters and their best-fit values are obtained using the $\chi^2$-analysis, which are consistent with the latest global neutrino oscillation experimental data. In our model, we have also explored the concept of NDBD. The value of effective Majorana neutrino mass $|m_{\beta\beta}|$ is well fitted within the sensitivity reach of the recent NDBD  experiments like KamLAND- Zen, GERDA and LEGEND-1k. The identification of NDBD, cosmological mass, and the leptonic CP-violation phase $\delta_{CP}$, which align with the most recent experimental information, will differentiate between various models of neutrino mass.

\begin{acknowledgments}
Animesh Barman acknowledges the financial support provided by the CSIR, New Delhi, India for Senior Research Fellowship (file no 09/796(0072)/2017-EMR-1). The research of Ng. K. Francis is funded by DST-SERB, India under Grant no. EMR/2015/001683.
\end{acknowledgments}

\bibliography{references}% Produces the bibliography via BibTeX.

\begin{thebibliography}{89}
\expandafter\ifx\csname natexlab\endcsname\relax\def\natexlab#1{#1}\fi
\expandafter\ifx\csname bibnamefont\endcsname\relax
  \def\bibnamefont#1{#1}\fi
\expandafter\ifx\csname bibfnamefont\endcsname\relax
  \def\bibfnamefont#1{#1}\fi
\expandafter\ifx\csname citenamefont\endcsname\relax
  \def\citenamefont#1{#1}\fi
\expandafter\ifx\csname url\endcsname\relax
  \def\url#1{\texttt{#1}}\fi
\expandafter\ifx\csname urlprefix\endcsname\relax\def\urlprefix{URL }\fi
\providecommand{\bibinfo}[2]{#2}
\providecommand{\eprint}[2][]{\url{#2}}

\bibitem[{\citenamefont{Maki et~al.}(1962)\citenamefont{Maki, Nakagawa, and
  Sakata}}]{Maki:1962mu}
\bibinfo{author}{\bibfnamefont{Z.}~\bibnamefont{Maki}},
  \bibinfo{author}{\bibfnamefont{M.}~\bibnamefont{Nakagawa}}, \bibnamefont{and}
  \bibinfo{author}{\bibfnamefont{S.}~\bibnamefont{Sakata}},
  \bibinfo{journal}{Prog. Theor. Phys.} \textbf{\bibinfo{volume}{28}},
  \bibinfo{pages}{870} (\bibinfo{year}{1962}).

\bibitem[{\citenamefont{Fusaoka and Katayama}(1972)}]{Fusaoka:1972hn}
\bibinfo{author}{\bibfnamefont{H.}~\bibnamefont{Fusaoka}} \bibnamefont{and}
  \bibinfo{author}{\bibfnamefont{Y.}~\bibnamefont{Katayama}},
  \bibinfo{journal}{Prog. Theor. Phys.} \textbf{\bibinfo{volume}{48}},
  \bibinfo{pages}{1753} (\bibinfo{year}{1972}).

\bibitem[{\citenamefont{Oberauer and von Feilitzsch}(1992)}]{Oberauer:1992vw}
\bibinfo{author}{\bibfnamefont{L.}~\bibnamefont{Oberauer}} \bibnamefont{and}
  \bibinfo{author}{\bibfnamefont{F.}~\bibnamefont{von Feilitzsch}},
  \bibinfo{journal}{Rept. Prog. Phys.} \textbf{\bibinfo{volume}{55}},
  \bibinfo{pages}{1093} (\bibinfo{year}{1992}).

\bibitem[{\citenamefont{Forero et~al.}(2012)\citenamefont{Forero, Tortola, and
  Valle}}]{Forero:2012faj}
\bibinfo{author}{\bibfnamefont{D.~V.} \bibnamefont{Forero}},
  \bibinfo{author}{\bibfnamefont{M.}~\bibnamefont{Tortola}}, \bibnamefont{and}
  \bibinfo{author}{\bibfnamefont{J.~W.~F.} \bibnamefont{Valle}},
  \bibinfo{journal}{Phys. Rev. D} \textbf{\bibinfo{volume}{86}},
  \bibinfo{pages}{073012} (\bibinfo{year}{2012}), \eprint{1205.4018}.

\bibitem[{\citenamefont{Aartsen et~al.}(2016)}]{Aartsen:2016psd}
\bibinfo{author}{\bibfnamefont{M.~G.} \bibnamefont{Aartsen}}
  \bibnamefont{et~al.}, \bibinfo{journal}{Nucl. Phys. B}
  \textbf{\bibinfo{volume}{908}}, \bibinfo{pages}{161} (\bibinfo{year}{2016}).

\bibitem[{\citenamefont{Gonzalez-Garcia
  et~al.}(2012)\citenamefont{Gonzalez-Garcia, Maltoni, Salvado, and
  Schwetz}}]{Gonzalez-Garcia:2012hef}
\bibinfo{author}{\bibfnamefont{M.~C.} \bibnamefont{Gonzalez-Garcia}},
  \bibinfo{author}{\bibfnamefont{M.}~\bibnamefont{Maltoni}},
  \bibinfo{author}{\bibfnamefont{J.}~\bibnamefont{Salvado}}, \bibnamefont{and}
  \bibinfo{author}{\bibfnamefont{T.}~\bibnamefont{Schwetz}},
  \bibinfo{journal}{JHEP} \textbf{\bibinfo{volume}{12}}, \bibinfo{pages}{123}
  (\bibinfo{year}{2012}), \eprint{1209.3023}.

\bibitem[{\citenamefont{Razzaghi}(2023)}]{Razzaghi:2023fsh}
\bibinfo{author}{\bibfnamefont{N.}~\bibnamefont{Razzaghi}},
  \bibinfo{journal}{Commun. Theor. Phys.} \textbf{\bibinfo{volume}{75}},
  \bibinfo{pages}{055204} (\bibinfo{year}{2023}).

\bibitem[{\citenamefont{Frank et~al.}(2022)\citenamefont{Frank, Majumdar,
  Poulose, Senapati, and Yajnik}}]{Frank:2022tbm}
\bibinfo{author}{\bibfnamefont{M.}~\bibnamefont{Frank}},
  \bibinfo{author}{\bibfnamefont{C.}~\bibnamefont{Majumdar}},
  \bibinfo{author}{\bibfnamefont{P.}~\bibnamefont{Poulose}},
  \bibinfo{author}{\bibfnamefont{S.}~\bibnamefont{Senapati}}, \bibnamefont{and}
  \bibinfo{author}{\bibfnamefont{U.~A.} \bibnamefont{Yajnik}},
  \bibinfo{journal}{JHEP} \textbf{\bibinfo{volume}{12}}, \bibinfo{pages}{032}
  (\bibinfo{year}{2022}), \eprint{2211.04286}.

\bibitem[{\citenamefont{Ahn et~al.}(2022)\citenamefont{Ahn, Kang, Ramos, and
  Tanimoto}}]{Ahn:2022ufs}
\bibinfo{author}{\bibfnamefont{Y.~H.} \bibnamefont{Ahn}},
  \bibinfo{author}{\bibfnamefont{S.~K.} \bibnamefont{Kang}},
  \bibinfo{author}{\bibfnamefont{R.}~\bibnamefont{Ramos}}, \bibnamefont{and}
  \bibinfo{author}{\bibfnamefont{M.}~\bibnamefont{Tanimoto}},
  \bibinfo{journal}{Phys. Rev. D} \textbf{\bibinfo{volume}{106}},
  \bibinfo{pages}{095002} (\bibinfo{year}{2022}), \eprint{2205.02796}.

\bibitem[{\citenamefont{Nath and Francis}(2021)}]{Nath:2018ywc}
\bibinfo{author}{\bibfnamefont{A.}~\bibnamefont{Nath}} \bibnamefont{and}
  \bibinfo{author}{\bibfnamefont{N.~K.} \bibnamefont{Francis}},
  \bibinfo{journal}{Int. J. Mod. Phys. A} \textbf{\bibinfo{volume}{36}},
  \bibinfo{pages}{2130008} (\bibinfo{year}{2021}), \eprint{1804.08467}.

\bibitem[{\citenamefont{Puyam et~al.}(2022)\citenamefont{Puyam, Singh, and
  Singh}}]{Puyam:2022mej}
\bibinfo{author}{\bibfnamefont{V.}~\bibnamefont{Puyam}},
  \bibinfo{author}{\bibfnamefont{S.~R.} \bibnamefont{Singh}}, \bibnamefont{and}
  \bibinfo{author}{\bibfnamefont{N.~N.} \bibnamefont{Singh}},
  \bibinfo{journal}{Nucl. Phys. B} \textbf{\bibinfo{volume}{983}},
  \bibinfo{pages}{115932} (\bibinfo{year}{2022}), \eprint{2204.10122}.

\bibitem[{\citenamefont{Kang and Dasgupta}(2019)}]{Kang:2019gyu}
\bibinfo{author}{\bibfnamefont{S.~K.} \bibnamefont{Kang}} \bibnamefont{and}
  \bibinfo{author}{\bibfnamefont{A.}~\bibnamefont{Dasgupta}},
  \bibinfo{journal}{PoS} \textbf{\bibinfo{volume}{ICHEP2018}},
  \bibinfo{pages}{410} (\bibinfo{year}{2019}).

\bibitem[{\citenamefont{Zhao}(2017)}]{Zhao:2017yvw}
\bibinfo{author}{\bibfnamefont{Z.-h.} \bibnamefont{Zhao}},
  \bibinfo{journal}{JHEP} \textbf{\bibinfo{volume}{09}}, \bibinfo{pages}{023}
  (\bibinfo{year}{2017}), \eprint{1703.04984}.

\bibitem[{\citenamefont{Buravov}(2017)}]{Buravov:2014dna}
\bibinfo{author}{\bibfnamefont{L.~I.} \bibnamefont{Buravov}},
  \bibinfo{journal}{J. Mod. Phys.} \textbf{\bibinfo{volume}{7}},
  \bibinfo{pages}{129} (\bibinfo{year}{2017}), \eprint{1502.00958}.

\bibitem[{\citenamefont{Nishi and Silva-Marcos}(2023)}]{Nishi:2023ebi}
\bibinfo{author}{\bibfnamefont{C.~C.} \bibnamefont{Nishi}} \bibnamefont{and}
  \bibinfo{author}{\bibfnamefont{J.~I.} \bibnamefont{Silva-Marcos}}
  (\bibinfo{year}{2023}), \eprint{2305.08980}.

\bibitem[{\citenamefont{Hagedorn et~al.}(2022)\citenamefont{Hagedorn, Kriewald,
  Orloff, and Teixeira}}]{Hagedorn:2021ldq}
\bibinfo{author}{\bibfnamefont{C.}~\bibnamefont{Hagedorn}},
  \bibinfo{author}{\bibfnamefont{J.}~\bibnamefont{Kriewald}},
  \bibinfo{author}{\bibfnamefont{J.}~\bibnamefont{Orloff}}, \bibnamefont{and}
  \bibinfo{author}{\bibfnamefont{A.~M.} \bibnamefont{Teixeira}},
  \bibinfo{journal}{Eur. Phys. J. C} \textbf{\bibinfo{volume}{82}},
  \bibinfo{pages}{194} (\bibinfo{year}{2022}), \eprint{2107.07537}.

\bibitem[{\citenamefont{Thapa et~al.}(2023)\citenamefont{Thapa, Barman, Bora,
  and Francis}}]{Thapa:2023fxu}
\bibinfo{author}{\bibfnamefont{B.}~\bibnamefont{Thapa}},
  \bibinfo{author}{\bibfnamefont{S.}~\bibnamefont{Barman}},
  \bibinfo{author}{\bibfnamefont{S.}~\bibnamefont{Bora}}, \bibnamefont{and}
  \bibinfo{author}{\bibfnamefont{N.~K.} \bibnamefont{Francis}}
  (\bibinfo{year}{2023}), \eprint{2305.09306}.

\bibitem[{\citenamefont{Bora et~al.}(2023)\citenamefont{Bora, Francis, Barman,
  and Thapa}}]{Bora:2023teg}
\bibinfo{author}{\bibfnamefont{H.}~\bibnamefont{Bora}},
  \bibinfo{author}{\bibfnamefont{N.~K.} \bibnamefont{Francis}},
  \bibinfo{author}{\bibfnamefont{A.}~\bibnamefont{Barman}}, \bibnamefont{and}
  \bibinfo{author}{\bibfnamefont{B.}~\bibnamefont{Thapa}}
  (\bibinfo{year}{2023}), \eprint{2305.08963}.

\bibitem[{\citenamefont{Loualidi}(2021)}]{Loualidi:2021qoq}
\bibinfo{author}{\bibfnamefont{M.~A.} \bibnamefont{Loualidi}}, in
  \emph{\bibinfo{booktitle}{{Beyond Standard Model: From Theory to
  Experiment}}} (\bibinfo{year}{2021}), \eprint{2104.13734}.

\bibitem[{\citenamefont{Khatun et~al.}(2020)\citenamefont{Khatun, Smetana, and
  \v{S}imkovic}}]{Khatun:2020biw}
\bibinfo{author}{\bibfnamefont{A.}~\bibnamefont{Khatun}},
  \bibinfo{author}{\bibfnamefont{A.}~\bibnamefont{Smetana}}, \bibnamefont{and}
  \bibinfo{author}{\bibfnamefont{F.}~\bibnamefont{\v{S}imkovic}},
  \bibinfo{journal}{Symmetry} \textbf{\bibinfo{volume}{12}},
  \bibinfo{pages}{1310} (\bibinfo{year}{2020}).

\bibitem[{\citenamefont{Okada and Orikasa}(2019)}]{Okada:2019lzv}
\bibinfo{author}{\bibfnamefont{H.}~\bibnamefont{Okada}} \bibnamefont{and}
  \bibinfo{author}{\bibfnamefont{Y.}~\bibnamefont{Orikasa}}
  (\bibinfo{year}{2019}), \eprint{1908.08409}.

\bibitem[{\citenamefont{Bhattacharyya and Datta}(2023)}]{Bhattacharyya:2022trp}
\bibinfo{author}{\bibfnamefont{S.}~\bibnamefont{Bhattacharyya}}
  \bibnamefont{and} \bibinfo{author}{\bibfnamefont{A.}~\bibnamefont{Datta}},
  \bibinfo{journal}{Nucl. Phys. B} \textbf{\bibinfo{volume}{991}},
  \bibinfo{pages}{116197} (\bibinfo{year}{2023}), \eprint{2206.13105}.

\bibitem[{\citenamefont{Boruah and Das}(2022)}]{Boruah:2021ktk}
\bibinfo{author}{\bibfnamefont{B.~B.} \bibnamefont{Boruah}} \bibnamefont{and}
  \bibinfo{author}{\bibfnamefont{M.~K.} \bibnamefont{Das}},
  \bibinfo{journal}{Int. J. Mod. Phys. A} \textbf{\bibinfo{volume}{37}},
  \bibinfo{pages}{2250026} (\bibinfo{year}{2022}), \eprint{2111.10341}.

\bibitem[{\citenamefont{Girardi}(2015)}]{Girardi:2015slc}
\bibinfo{author}{\bibfnamefont{I.}~\bibnamefont{Girardi}}, Ph.D. thesis,
  \bibinfo{school}{SISSA, Trieste} (\bibinfo{year}{2015}).

\bibitem[{\citenamefont{Mishra et~al.}(2023)\citenamefont{Mishra, Behera, and
  Mohanta}}]{Mishra:2023cjc}
\bibinfo{author}{\bibfnamefont{P.}~\bibnamefont{Mishra}},
  \bibinfo{author}{\bibfnamefont{M.~K.} \bibnamefont{Behera}},
  \bibnamefont{and} \bibinfo{author}{\bibfnamefont{R.}~\bibnamefont{Mohanta}},
  \bibinfo{journal}{Phys. Rev. D} \textbf{\bibinfo{volume}{107}},
  \bibinfo{pages}{115004} (\bibinfo{year}{2023}), \eprint{2302.00494}.

\bibitem[{\citenamefont{Ahn et~al.}(2012)}]{RENO:2012mkc}
\bibinfo{author}{\bibfnamefont{J.~K.} \bibnamefont{Ahn}} \bibnamefont{et~al.}
  (\bibinfo{collaboration}{RENO}), \bibinfo{journal}{Phys. Rev. Lett.}
  \textbf{\bibinfo{volume}{108}}, \bibinfo{pages}{191802}
  (\bibinfo{year}{2012}), \eprint{1204.0626}.

\bibitem[{\citenamefont{Abe et~al.}(2012)}]{DoubleChooz:2011ymz}
\bibinfo{author}{\bibfnamefont{Y.}~\bibnamefont{Abe}} \bibnamefont{et~al.}
  (\bibinfo{collaboration}{Double Chooz}), \bibinfo{journal}{Phys. Rev. Lett.}
  \textbf{\bibinfo{volume}{108}}, \bibinfo{pages}{131801}
  (\bibinfo{year}{2012}), \eprint{1112.6353}.

\bibitem[{\citenamefont{An et~al.}(2012)}]{DayaBay:2012fng}
\bibinfo{author}{\bibfnamefont{F.~P.} \bibnamefont{An}} \bibnamefont{et~al.}
  (\bibinfo{collaboration}{Daya Bay}), \bibinfo{journal}{Phys. Rev. Lett.}
  \textbf{\bibinfo{volume}{108}}, \bibinfo{pages}{171803}
  (\bibinfo{year}{2012}), \eprint{1203.1669}.

\bibitem[{\citenamefont{Adamson et~al.}(2011)}]{MINOS:2011amj}
\bibinfo{author}{\bibfnamefont{P.}~\bibnamefont{Adamson}} \bibnamefont{et~al.}
  (\bibinfo{collaboration}{MINOS}), \bibinfo{journal}{Phys. Rev. Lett.}
  \textbf{\bibinfo{volume}{107}}, \bibinfo{pages}{181802}
  (\bibinfo{year}{2011}), \eprint{1108.0015}.

\bibitem[{\citenamefont{Abe et~al.}(2011)}]{T2K:2011ypd}
\bibinfo{author}{\bibfnamefont{K.}~\bibnamefont{Abe}} \bibnamefont{et~al.}
  (\bibinfo{collaboration}{T2K}), \bibinfo{journal}{Phys. Rev. Lett.}
  \textbf{\bibinfo{volume}{107}}, \bibinfo{pages}{041801}
  (\bibinfo{year}{2011}), \eprint{1106.2822}.

\bibitem[{\citenamefont{Blennow and Schwetz}(2012)}]{blennow2012identifying}
\bibinfo{author}{\bibfnamefont{M.}~\bibnamefont{Blennow}} \bibnamefont{and}
  \bibinfo{author}{\bibfnamefont{T.}~\bibnamefont{Schwetz}},
  \bibinfo{journal}{Journal of High Energy Physics}
  \textbf{\bibinfo{volume}{2012}}, \bibinfo{pages}{1} (\bibinfo{year}{2012}).

\bibitem[{\citenamefont{Ghosh et~al.}(2013)\citenamefont{Ghosh, Thakore, and
  Choubey}}]{ghosh2013determining}
\bibinfo{author}{\bibfnamefont{A.}~\bibnamefont{Ghosh}},
  \bibinfo{author}{\bibfnamefont{T.}~\bibnamefont{Thakore}}, \bibnamefont{and}
  \bibinfo{author}{\bibfnamefont{S.}~\bibnamefont{Choubey}},
  \bibinfo{journal}{Journal of High Energy Physics}
  \textbf{\bibinfo{volume}{2013}}, \bibinfo{pages}{1} (\bibinfo{year}{2013}).

\bibitem[{\citenamefont{Ahmed et~al.}(2017)}]{ICAL:2015stm}
\bibinfo{author}{\bibfnamefont{S.}~\bibnamefont{Ahmed}} \bibnamefont{et~al.}
  (\bibinfo{collaboration}{ICAL}), \bibinfo{journal}{Pramana}
  \textbf{\bibinfo{volume}{88}}, \bibinfo{pages}{79} (\bibinfo{year}{2017}),
  \eprint{1505.07380}.

\bibitem[{\citenamefont{Ribordy and Smirnov}(2013)}]{Ribordy:2013xea}
\bibinfo{author}{\bibfnamefont{M.}~\bibnamefont{Ribordy}} \bibnamefont{and}
  \bibinfo{author}{\bibfnamefont{A.~Y.} \bibnamefont{Smirnov}},
  \bibinfo{journal}{Phys. Rev. D} \textbf{\bibinfo{volume}{87}},
  \bibinfo{pages}{113007} (\bibinfo{year}{2013}), \eprint{1303.0758}.

\bibitem[{\citenamefont{Quinn}(2018)}]{Quinn:2018zvt}
\bibinfo{author}{\bibfnamefont{L.}~\bibnamefont{Quinn}}, Ph.D. thesis,
  \bibinfo{school}{Aix-Marseille U.} (\bibinfo{year}{2018}).

\bibitem[{\citenamefont{Jia et~al.}(2019)\citenamefont{Jia, Wang, and
  Zhou}}]{Jia:2017oar}
\bibinfo{author}{\bibfnamefont{J.}~\bibnamefont{Jia}},
  \bibinfo{author}{\bibfnamefont{Y.}~\bibnamefont{Wang}}, \bibnamefont{and}
  \bibinfo{author}{\bibfnamefont{S.}~\bibnamefont{Zhou}},
  \bibinfo{journal}{Chin. Phys. C} \textbf{\bibinfo{volume}{43}},
  \bibinfo{pages}{095102} (\bibinfo{year}{2019}), \eprint{1709.09453}.

\bibitem[{\citenamefont{Winter}(2013)}]{Winter:2013ema}
\bibinfo{author}{\bibfnamefont{W.}~\bibnamefont{Winter}},
  \bibinfo{journal}{Phys. Rev. D} \textbf{\bibinfo{volume}{88}},
  \bibinfo{pages}{013013} (\bibinfo{year}{2013}), \eprint{1305.5539}.

\bibitem[{\citenamefont{Agarwalla et~al.}(2014)\citenamefont{Agarwalla,
  Agostino, Aittola, Alekou, Andrieu, Antoniou, Asfandiyarov, Autiero,
  B{\'e}sida, Balik et~al.}}]{agarwalla2014lbno}
\bibinfo{author}{\bibfnamefont{S.}~\bibnamefont{Agarwalla}},
  \bibinfo{author}{\bibfnamefont{L.}~\bibnamefont{Agostino}},
  \bibinfo{author}{\bibfnamefont{M.}~\bibnamefont{Aittola}},
  \bibinfo{author}{\bibfnamefont{A.}~\bibnamefont{Alekou}},
  \bibinfo{author}{\bibfnamefont{B.}~\bibnamefont{Andrieu}},
  \bibinfo{author}{\bibfnamefont{F.}~\bibnamefont{Antoniou}},
  \bibinfo{author}{\bibfnamefont{R.}~\bibnamefont{Asfandiyarov}},
  \bibinfo{author}{\bibfnamefont{D.}~\bibnamefont{Autiero}},
  \bibinfo{author}{\bibfnamefont{O.}~\bibnamefont{B{\'e}sida}},
  \bibinfo{author}{\bibfnamefont{A.}~\bibnamefont{Balik}},
  \bibnamefont{et~al.}, \bibinfo{journal}{arXiv preprint arXiv:1412.0804}
  (\bibinfo{year}{2014}).

\bibitem[{\citenamefont{Agarwalla}(2013)}]{agarwalla2013neutrino}
\bibinfo{author}{\bibfnamefont{S.~K.} \bibnamefont{Agarwalla}},
  \bibinfo{journal}{Nuclear Physics B-Proceedings Supplements}
  \textbf{\bibinfo{volume}{237}}, \bibinfo{pages}{196} (\bibinfo{year}{2013}).

\bibitem[{\citenamefont{Esteban et~al.}(2020)\citenamefont{Esteban,
  Gonz{\'a}lez-Garc{\'\i}a, Maltoni, Schwetz, and Zhou}}]{esteban2020fate}
\bibinfo{author}{\bibfnamefont{I.}~\bibnamefont{Esteban}},
  \bibinfo{author}{\bibfnamefont{M.~C.}
  \bibnamefont{Gonz{\'a}lez-Garc{\'\i}a}},
  \bibinfo{author}{\bibfnamefont{M.}~\bibnamefont{Maltoni}},
  \bibinfo{author}{\bibfnamefont{T.}~\bibnamefont{Schwetz}}, \bibnamefont{and}
  \bibinfo{author}{\bibfnamefont{A.}~\bibnamefont{Zhou}},
  \bibinfo{journal}{Journal of High Energy Physics}
  \textbf{\bibinfo{volume}{2020}}, \bibinfo{pages}{1} (\bibinfo{year}{2020}).

\bibitem[{\citenamefont{Minkowski}(1977)}]{minkowski1977mu}
\bibinfo{author}{\bibfnamefont{P.}~\bibnamefont{Minkowski}},
  \bibinfo{journal}{Physics Letters B} \textbf{\bibinfo{volume}{67}},
  \bibinfo{pages}{421} (\bibinfo{year}{1977}).

\bibitem[{\citenamefont{Yanagida}(1979)}]{yanagida1979proc}
\bibinfo{author}{\bibfnamefont{T.}~\bibnamefont{Yanagida}},
  \bibinfo{journal}{KEK Report No. 79-18} \textbf{\bibinfo{volume}{95}}
  (\bibinfo{year}{1979}).

\bibitem[{\citenamefont{Mohapatra and
  Senjanovi{\'c}}(1980)}]{mohapatra1980neutrino}
\bibinfo{author}{\bibfnamefont{R.~N.} \bibnamefont{Mohapatra}}
  \bibnamefont{and}
  \bibinfo{author}{\bibfnamefont{G.}~\bibnamefont{Senjanovi{\'c}}},
  \bibinfo{journal}{Physical Review Letters} \textbf{\bibinfo{volume}{44}},
  \bibinfo{pages}{912} (\bibinfo{year}{1980}).

\bibitem[{\citenamefont{Gell-Mann et~al.}(1979)\citenamefont{Gell-Mann, Ramond,
  and Slansky}}]{gell1979supergravity}
\bibinfo{author}{\bibfnamefont{M.}~\bibnamefont{Gell-Mann}},
  \bibinfo{author}{\bibfnamefont{P.}~\bibnamefont{Ramond}}, \bibnamefont{and}
  \bibinfo{author}{\bibfnamefont{R.}~\bibnamefont{Slansky}},
  \bibinfo{journal}{Amsterdam: North-Holland) p}
  \textbf{\bibinfo{volume}{315}}, \bibinfo{pages}{79} (\bibinfo{year}{1979}).

\bibitem[{\citenamefont{Achiman and Stech}(1978)}]{achiman1978quark}
\bibinfo{author}{\bibfnamefont{Y.}~\bibnamefont{Achiman}} \bibnamefont{and}
  \bibinfo{author}{\bibfnamefont{B.}~\bibnamefont{Stech}},
  \bibinfo{journal}{Physics Letters B} \textbf{\bibinfo{volume}{77}},
  \bibinfo{pages}{389} (\bibinfo{year}{1978}).

\bibitem[{\citenamefont{Ma}(1999)}]{ma1999supersymmetry}
\bibinfo{author}{\bibfnamefont{E.}~\bibnamefont{Ma}}, \bibinfo{journal}{arXiv
  preprint hep-ph/9902450}  (\bibinfo{year}{1999}).

\bibitem[{\citenamefont{Csaki}(1996)}]{Csaki:1996ks}
\bibinfo{author}{\bibfnamefont{C.}~\bibnamefont{Csaki}}, \bibinfo{journal}{Mod.
  Phys. Lett. A} \textbf{\bibinfo{volume}{11}}, \bibinfo{pages}{599}
  (\bibinfo{year}{1996}), \eprint{hep-ph/9606414}.

\bibitem[{\citenamefont{Kang}(2021)}]{kang2021low}
\bibinfo{author}{\bibfnamefont{S.~K.} \bibnamefont{Kang}},
  \bibinfo{journal}{Journal of the Korean Physical Society}
  \textbf{\bibinfo{volume}{78}}, \bibinfo{pages}{743} (\bibinfo{year}{2021}).

\bibitem[{\citenamefont{Hue et~al.}(2022)\citenamefont{Hue, Phan, Nguyen, Long,
  and Hung}}]{Hue:2021zyw}
\bibinfo{author}{\bibfnamefont{L.~T.} \bibnamefont{Hue}},
  \bibinfo{author}{\bibfnamefont{K.~H.} \bibnamefont{Phan}},
  \bibinfo{author}{\bibfnamefont{T.~P.} \bibnamefont{Nguyen}},
  \bibinfo{author}{\bibfnamefont{H.~N.} \bibnamefont{Long}}, \bibnamefont{and}
  \bibinfo{author}{\bibfnamefont{H.~T.} \bibnamefont{Hung}},
  \bibinfo{journal}{Eur. Phys. J. C} \textbf{\bibinfo{volume}{82}},
  \bibinfo{pages}{722} (\bibinfo{year}{2022}), \eprint{2109.06089}.

\bibitem[{\citenamefont{Ellwanger et~al.}(2010)\citenamefont{Ellwanger,
  Hugonie, and Teixeira}}]{Ellwanger:2009dp}
\bibinfo{author}{\bibfnamefont{U.}~\bibnamefont{Ellwanger}},
  \bibinfo{author}{\bibfnamefont{C.}~\bibnamefont{Hugonie}}, \bibnamefont{and}
  \bibinfo{author}{\bibfnamefont{A.~M.} \bibnamefont{Teixeira}},
  \bibinfo{journal}{Phys. Rept.} \textbf{\bibinfo{volume}{496}},
  \bibinfo{pages}{1} (\bibinfo{year}{2010}), \eprint{0910.1785}.

\bibitem[{\citenamefont{Ibanez and Uranga}(2012)}]{ibanez2012string}
\bibinfo{author}{\bibfnamefont{L.~E.} \bibnamefont{Ibanez}} \bibnamefont{and}
  \bibinfo{author}{\bibfnamefont{A.~M.} \bibnamefont{Uranga}},
  \emph{\bibinfo{title}{String theory and particle physics: An introduction to
  string phenomenology}} (\bibinfo{publisher}{Cambridge University Press},
  \bibinfo{year}{2012}).

\bibitem[{\citenamefont{Arkani-Hamed et~al.}(2001)\citenamefont{Arkani-Hamed,
  Dimopoulos, Dvali, and March-Russell}}]{Arkani-Hamed:1998wuz}
\bibinfo{author}{\bibfnamefont{N.}~\bibnamefont{Arkani-Hamed}},
  \bibinfo{author}{\bibfnamefont{S.}~\bibnamefont{Dimopoulos}},
  \bibinfo{author}{\bibfnamefont{G.~R.} \bibnamefont{Dvali}}, \bibnamefont{and}
  \bibinfo{author}{\bibfnamefont{J.}~\bibnamefont{March-Russell}},
  \bibinfo{journal}{Phys. Rev. D} \textbf{\bibinfo{volume}{65}},
  \bibinfo{pages}{024032} (\bibinfo{year}{2001}), \eprint{hep-ph/9811448}.

\bibitem[{\citenamefont{Thapa and Francis}(2023)}]{Thapa:2022fhv}
\bibinfo{author}{\bibfnamefont{B.}~\bibnamefont{Thapa}} \bibnamefont{and}
  \bibinfo{author}{\bibfnamefont{N.~K.} \bibnamefont{Francis}},
  \bibinfo{journal}{Nucl. Phys. B} \textbf{\bibinfo{volume}{986}},
  \bibinfo{pages}{116054} (\bibinfo{year}{2023}), \eprint{2209.06263}.

\bibitem[{\citenamefont{Ma}(2006{\natexlab{a}})}]{Ma:2006km}
\bibinfo{author}{\bibfnamefont{E.}~\bibnamefont{Ma}}, \bibinfo{journal}{Phys.
  Rev. D} \textbf{\bibinfo{volume}{73}}, \bibinfo{pages}{077301}
  (\bibinfo{year}{2006}{\natexlab{a}}), \eprint{hep-ph/0601225}.

\bibitem[{\citenamefont{King and Luhn}(2013)}]{King:2013eh}
\bibinfo{author}{\bibfnamefont{S.~F.} \bibnamefont{King}} \bibnamefont{and}
  \bibinfo{author}{\bibfnamefont{C.}~\bibnamefont{Luhn}},
  \bibinfo{journal}{Rept. Prog. Phys.} \textbf{\bibinfo{volume}{76}},
  \bibinfo{pages}{056201} (\bibinfo{year}{2013}), \eprint{1301.1340}.

\bibitem[{\citenamefont{Ishimori et~al.}(2010)\citenamefont{Ishimori,
  Kobayashi, Ohki, Shimizu, Okada, and Tanimoto}}]{Ishimori:2010au}
\bibinfo{author}{\bibfnamefont{H.}~\bibnamefont{Ishimori}},
  \bibinfo{author}{\bibfnamefont{T.}~\bibnamefont{Kobayashi}},
  \bibinfo{author}{\bibfnamefont{H.}~\bibnamefont{Ohki}},
  \bibinfo{author}{\bibfnamefont{Y.}~\bibnamefont{Shimizu}},
  \bibinfo{author}{\bibfnamefont{H.}~\bibnamefont{Okada}}, \bibnamefont{and}
  \bibinfo{author}{\bibfnamefont{M.}~\bibnamefont{Tanimoto}},
  \bibinfo{journal}{Prog. Theor. Phys. Suppl.} \textbf{\bibinfo{volume}{183}},
  \bibinfo{pages}{1} (\bibinfo{year}{2010}), \eprint{1003.3552}.

\bibitem[{\citenamefont{Nguyen et~al.}(2022)\citenamefont{Nguyen, Thuc, Si,
  Hong, and Hue}}]{Nguyen:2020ehj}
\bibinfo{author}{\bibfnamefont{T.~P.} \bibnamefont{Nguyen}},
  \bibinfo{author}{\bibfnamefont{T.~T.} \bibnamefont{Thuc}},
  \bibinfo{author}{\bibfnamefont{D.~T.} \bibnamefont{Si}},
  \bibinfo{author}{\bibfnamefont{T.~T.} \bibnamefont{Hong}}, \bibnamefont{and}
  \bibinfo{author}{\bibfnamefont{L.~T.} \bibnamefont{Hue}},
  \bibinfo{journal}{PTEP} \textbf{\bibinfo{volume}{2022}},
  \bibinfo{pages}{023B01} (\bibinfo{year}{2022}), \eprint{2011.12181}.

\bibitem[{\citenamefont{Ganguly et~al.}(2022)\citenamefont{Ganguly, Gluza, and
  Karmakar}}]{Ganguly:2022qxj}
\bibinfo{author}{\bibfnamefont{J.}~\bibnamefont{Ganguly}},
  \bibinfo{author}{\bibfnamefont{J.}~\bibnamefont{Gluza}}, \bibnamefont{and}
  \bibinfo{author}{\bibfnamefont{B.}~\bibnamefont{Karmakar}},
  \bibinfo{journal}{JHEP} \textbf{\bibinfo{volume}{11}}, \bibinfo{pages}{074}
  (\bibinfo{year}{2022}), \eprint{2209.08610}.

\bibitem[{\citenamefont{Behera et~al.}(2022)\citenamefont{Behera, Mishra,
  Singirala, and Mohanta}}]{Behera:2020sfe}
\bibinfo{author}{\bibfnamefont{M.~K.} \bibnamefont{Behera}},
  \bibinfo{author}{\bibfnamefont{S.}~\bibnamefont{Mishra}},
  \bibinfo{author}{\bibfnamefont{S.}~\bibnamefont{Singirala}},
  \bibnamefont{and} \bibinfo{author}{\bibfnamefont{R.}~\bibnamefont{Mohanta}},
  \bibinfo{journal}{Phys. Dark Univ.} \textbf{\bibinfo{volume}{36}},
  \bibinfo{pages}{101027} (\bibinfo{year}{2022}), \eprint{2007.00545}.

\bibitem[{\citenamefont{Vien and Long}(2023)}]{Vien:2022cxr}
\bibinfo{author}{\bibfnamefont{V.~V.} \bibnamefont{Vien}} \bibnamefont{and}
  \bibinfo{author}{\bibfnamefont{H.~N.} \bibnamefont{Long}},
  \bibinfo{journal}{Phys. Scripta} \textbf{\bibinfo{volume}{98}},
  \bibinfo{pages}{015301} (\bibinfo{year}{2023}), \eprint{2302.03195}.

\bibitem[{\citenamefont{Ma}(2006{\natexlab{b}})}]{Ma:2005qf}
\bibinfo{author}{\bibfnamefont{E.}~\bibnamefont{Ma}}, \bibinfo{journal}{Phys.
  Rev. D} \textbf{\bibinfo{volume}{73}}, \bibinfo{pages}{057304}
  (\bibinfo{year}{2006}{\natexlab{b}}), \eprint{hep-ph/0511133}.

\bibitem[{\citenamefont{Ma}(2016)}]{Ma:2015fpa}
\bibinfo{author}{\bibfnamefont{E.}~\bibnamefont{Ma}}, \bibinfo{journal}{Phys.
  Lett. B} \textbf{\bibinfo{volume}{752}}, \bibinfo{pages}{198}
  (\bibinfo{year}{2016}), \eprint{1510.02501}.

\bibitem[{\citenamefont{Ma}(2004)}]{ma2004non}
\bibinfo{author}{\bibfnamefont{E.}~\bibnamefont{Ma}}, \bibinfo{journal}{arXiv
  preprint hep-ph/0405152}  (\bibinfo{year}{2004}).

\bibitem[{\citenamefont{Bazzocchi and Merlo}(2013)}]{Bazzocchi:2012st}
\bibinfo{author}{\bibfnamefont{F.}~\bibnamefont{Bazzocchi}} \bibnamefont{and}
  \bibinfo{author}{\bibfnamefont{L.}~\bibnamefont{Merlo}},
  \bibinfo{journal}{Fortsch. Phys.} \textbf{\bibinfo{volume}{61}},
  \bibinfo{pages}{571} (\bibinfo{year}{2013}), \eprint{1205.5135}.

\bibitem[{\citenamefont{Vien and Long}(2021)}]{Vien:2020aya}
\bibinfo{author}{\bibfnamefont{V.~V.} \bibnamefont{Vien}} \bibnamefont{and}
  \bibinfo{author}{\bibfnamefont{H.~N.} \bibnamefont{Long}},
  \bibinfo{journal}{Chin. Phys. C} \textbf{\bibinfo{volume}{45}},
  \bibinfo{pages}{043112} (\bibinfo{year}{2021}), \eprint{2012.01715}.

\bibitem[{\citenamefont{Thapa and Francis}(2021)}]{Thapa:2021ehj}
\bibinfo{author}{\bibfnamefont{B.}~\bibnamefont{Thapa}} \bibnamefont{and}
  \bibinfo{author}{\bibfnamefont{N.~K.} \bibnamefont{Francis}},
  \bibinfo{journal}{Eur. Phys. J. C} \textbf{\bibinfo{volume}{81}},
  \bibinfo{pages}{1061} (\bibinfo{year}{2021}), \eprint{2107.02074}.

\bibitem[{\citenamefont{Ma}(2005)}]{Ma:2005mw}
\bibinfo{author}{\bibfnamefont{E.}~\bibnamefont{Ma}}, \bibinfo{journal}{Mod.
  Phys. Lett. A} \textbf{\bibinfo{volume}{20}}, \bibinfo{pages}{2601}
  (\bibinfo{year}{2005}), \eprint{hep-ph/0508099}.

\bibitem[{\citenamefont{Dev et~al.}(2015)\citenamefont{Dev, Ramadevi, and
  Sankar}}]{dev2015non}
\bibinfo{author}{\bibfnamefont{A.}~\bibnamefont{Dev}},
  \bibinfo{author}{\bibfnamefont{P.}~\bibnamefont{Ramadevi}}, \bibnamefont{and}
  \bibinfo{author}{\bibfnamefont{S.~U.} \bibnamefont{Sankar}},
  \bibinfo{journal}{Journal of High Energy Physics}
  \textbf{\bibinfo{volume}{2015}}, \bibinfo{pages}{1} (\bibinfo{year}{2015}).

\bibitem[{\citenamefont{Grossman and Ng}(2015)}]{Grossman:2014oqa}
\bibinfo{author}{\bibfnamefont{Y.}~\bibnamefont{Grossman}} \bibnamefont{and}
  \bibinfo{author}{\bibfnamefont{W.~H.} \bibnamefont{Ng}},
  \bibinfo{journal}{Phys. Rev. D} \textbf{\bibinfo{volume}{91}},
  \bibinfo{pages}{073005} (\bibinfo{year}{2015}), \eprint{1404.1413}.

\bibitem[{\citenamefont{Ma}(2008)}]{Ma:2007wu}
\bibinfo{author}{\bibfnamefont{E.}~\bibnamefont{Ma}}, \bibinfo{journal}{Phys.
  Lett. B} \textbf{\bibinfo{volume}{660}}, \bibinfo{pages}{505}
  (\bibinfo{year}{2008}), \eprint{0709.0507}.

\bibitem[{\citenamefont{Hern{\'a}ndez et~al.}(2016)\citenamefont{Hern{\'a}ndez,
  Long, and Vien}}]{hernandez20163}
\bibinfo{author}{\bibfnamefont{A.~C.} \bibnamefont{Hern{\'a}ndez}},
  \bibinfo{author}{\bibfnamefont{H.}~\bibnamefont{Long}}, \bibnamefont{and}
  \bibinfo{author}{\bibfnamefont{V.}~\bibnamefont{Vien}}, \bibinfo{journal}{The
  European Physical Journal C} \textbf{\bibinfo{volume}{76}},
  \bibinfo{pages}{1} (\bibinfo{year}{2016}).

\bibitem[{\citenamefont{de~Medeiros~Varzielas
  et~al.}(2007)\citenamefont{de~Medeiros~Varzielas, King, and
  Ross}}]{de2007neutrino}
\bibinfo{author}{\bibfnamefont{I.}~\bibnamefont{de~Medeiros~Varzielas}},
  \bibinfo{author}{\bibfnamefont{S.}~\bibnamefont{King}}, \bibnamefont{and}
  \bibinfo{author}{\bibfnamefont{G.}~\bibnamefont{Ross}},
  \bibinfo{journal}{Physics Letters B} \textbf{\bibinfo{volume}{648}},
  \bibinfo{pages}{201} (\bibinfo{year}{2007}).

\bibitem[{\citenamefont{C\'arcamo~Hern\'andez and
  de~Medeiros~Varzielas}(2020)}]{CarcamoHernandez:2020udg}
\bibinfo{author}{\bibfnamefont{A.~E.} \bibnamefont{C\'arcamo~Hern\'andez}}
  \bibnamefont{and}
  \bibinfo{author}{\bibfnamefont{I.}~\bibnamefont{de~Medeiros~Varzielas}},
  \bibinfo{journal}{Phys. Lett. B} \textbf{\bibinfo{volume}{806}},
  \bibinfo{pages}{135491} (\bibinfo{year}{2020}), \eprint{2003.01134}.

\bibitem[{\citenamefont{Ishimori et~al.}(2009)\citenamefont{Ishimori,
  Kobayashi, Okada, Shimizu, and Tanimoto}}]{ishimori2009lepton}
\bibinfo{author}{\bibfnamefont{H.}~\bibnamefont{Ishimori}},
  \bibinfo{author}{\bibfnamefont{T.}~\bibnamefont{Kobayashi}},
  \bibinfo{author}{\bibfnamefont{H.}~\bibnamefont{Okada}},
  \bibinfo{author}{\bibfnamefont{Y.}~\bibnamefont{Shimizu}}, \bibnamefont{and}
  \bibinfo{author}{\bibfnamefont{M.}~\bibnamefont{Tanimoto}},
  \bibinfo{journal}{Journal of High Energy Physics}
  \textbf{\bibinfo{volume}{2009}}, \bibinfo{pages}{011} (\bibinfo{year}{2009}).

\bibitem[{\citenamefont{Carballo-P{\'e}rez
  et~al.}(2016)\citenamefont{Carballo-P{\'e}rez, Peinado, and
  Ramos-S{\'a}nchez}}]{carballo2016delta}
\bibinfo{author}{\bibfnamefont{B.}~\bibnamefont{Carballo-P{\'e}rez}},
  \bibinfo{author}{\bibfnamefont{E.}~\bibnamefont{Peinado}}, \bibnamefont{and}
  \bibinfo{author}{\bibfnamefont{S.}~\bibnamefont{Ramos-S{\'a}nchez}},
  \bibinfo{journal}{Journal of High Energy Physics}
  \textbf{\bibinfo{volume}{2016}}, \bibinfo{pages}{1} (\bibinfo{year}{2016}).

\bibitem[{\citenamefont{Liu et~al.}(2016)\citenamefont{Liu, Zhang, and
  Zhou}}]{Liu:2016oph}
\bibinfo{author}{\bibfnamefont{J.-H.} \bibnamefont{Liu}},
  \bibinfo{author}{\bibfnamefont{J.}~\bibnamefont{Zhang}}, \bibnamefont{and}
  \bibinfo{author}{\bibfnamefont{S.}~\bibnamefont{Zhou}},
  \bibinfo{journal}{Phys. Lett. B} \textbf{\bibinfo{volume}{760}},
  \bibinfo{pages}{571} (\bibinfo{year}{2016}), \eprint{1606.04886}.

\bibitem[{\citenamefont{Altarelli and Feruglio}(2006)}]{Altarelli:2005yx}
\bibinfo{author}{\bibfnamefont{G.}~\bibnamefont{Altarelli}} \bibnamefont{and}
  \bibinfo{author}{\bibfnamefont{F.}~\bibnamefont{Feruglio}},
  \bibinfo{journal}{Nucl. Phys. B} \textbf{\bibinfo{volume}{741}},
  \bibinfo{pages}{215} (\bibinfo{year}{2006}), \eprint{hep-ph/0512103}.

\bibitem[{\citenamefont{Altarelli and Feruglio}(2010)}]{Altarelli:2010gt}
\bibinfo{author}{\bibfnamefont{G.}~\bibnamefont{Altarelli}} \bibnamefont{and}
  \bibinfo{author}{\bibfnamefont{F.}~\bibnamefont{Feruglio}},
  \bibinfo{journal}{Rev. Mod. Phys.} \textbf{\bibinfo{volume}{82}},
  \bibinfo{pages}{2701} (\bibinfo{year}{2010}), \eprint{1002.0211}.

\bibitem[{\citenamefont{Borah and Karmakar}(2018)}]{Borah:2017dmk}
\bibinfo{author}{\bibfnamefont{D.}~\bibnamefont{Borah}} \bibnamefont{and}
  \bibinfo{author}{\bibfnamefont{B.}~\bibnamefont{Karmakar}},
  \bibinfo{journal}{Phys. Lett. B} \textbf{\bibinfo{volume}{780}},
  \bibinfo{pages}{461} (\bibinfo{year}{2018}), \eprint{1712.06407}.

\bibitem[{\citenamefont{Crivellin et~al.}(2022)\citenamefont{Crivellin, Kirk,
  and Manzari}}]{Crivellin:2022cve}
\bibinfo{author}{\bibfnamefont{A.}~\bibnamefont{Crivellin}},
  \bibinfo{author}{\bibfnamefont{F.}~\bibnamefont{Kirk}}, \bibnamefont{and}
  \bibinfo{author}{\bibfnamefont{C.~A.} \bibnamefont{Manzari}},
  \bibinfo{journal}{JHEP} \textbf{\bibinfo{volume}{12}}, \bibinfo{pages}{031}
  (\bibinfo{year}{2022}), \eprint{2208.00020}.

\bibitem[{\citenamefont{Kobayashi et~al.}(2022)\citenamefont{Kobayashi, Ohki,
  Okada, Shimizu, and Tanimoto}}]{Kobayashi:2022moq}
\bibinfo{author}{\bibfnamefont{T.}~\bibnamefont{Kobayashi}},
  \bibinfo{author}{\bibfnamefont{H.}~\bibnamefont{Ohki}},
  \bibinfo{author}{\bibfnamefont{H.}~\bibnamefont{Okada}},
  \bibinfo{author}{\bibfnamefont{Y.}~\bibnamefont{Shimizu}}, \bibnamefont{and}
  \bibinfo{author}{\bibfnamefont{M.}~\bibnamefont{Tanimoto}},
  \emph{\bibinfo{title}{{An Introduction to Non-Abelian Discrete Symmetries for
  Particle Physicists}}} (\bibinfo{publisher}{Springer}, \bibinfo{year}{2022}).

\bibitem[{\citenamefont{Barman et~al.}(2023)\citenamefont{Barman, Francis,
  Thapa, and Nath}}]{Barman:2022hyq}
\bibinfo{author}{\bibfnamefont{A.}~\bibnamefont{Barman}},
  \bibinfo{author}{\bibfnamefont{N.~K.} \bibnamefont{Francis}},
  \bibinfo{author}{\bibfnamefont{B.}~\bibnamefont{Thapa}}, \bibnamefont{and}
  \bibinfo{author}{\bibfnamefont{A.}~\bibnamefont{Nath}},
  \bibinfo{journal}{Int. J. Mod. Phys. A} \textbf{\bibinfo{volume}{38}},
  \bibinfo{pages}{2350012} (\bibinfo{year}{2023}), \eprint{2203.05536}.

\bibitem[{\citenamefont{Brahmachari et~al.}(2008)\citenamefont{Brahmachari,
  Choubey, and Mitra}}]{Brahmachari:2008fn}
\bibinfo{author}{\bibfnamefont{B.}~\bibnamefont{Brahmachari}},
  \bibinfo{author}{\bibfnamefont{S.}~\bibnamefont{Choubey}}, \bibnamefont{and}
  \bibinfo{author}{\bibfnamefont{M.}~\bibnamefont{Mitra}},
  \bibinfo{journal}{Phys. Rev. D} \textbf{\bibinfo{volume}{77}},
  \bibinfo{pages}{073008} (\bibinfo{year}{2008}), \bibinfo{note}{[Erratum:
  Phys.Rev.D 77, 119901 (2008)]}, \eprint{0801.3554}.

\bibitem[{\citenamefont{Gando}(2020)}]{Gando:2020cxo}
\bibinfo{author}{\bibfnamefont{Y.}~\bibnamefont{Gando}}
  (\bibinfo{collaboration}{KamLAND-Zen}), \bibinfo{journal}{J. Phys. Conf.
  Ser.} \textbf{\bibinfo{volume}{1468}}, \bibinfo{pages}{012142}
  (\bibinfo{year}{2020}).

\bibitem[{\citenamefont{Biller}(2021)}]{Biller:2021bqx}
\bibinfo{author}{\bibfnamefont{S.~D.} \bibnamefont{Biller}},
  \bibinfo{journal}{Phys. Rev. D} \textbf{\bibinfo{volume}{104}},
  \bibinfo{pages}{012002} (\bibinfo{year}{2021}), \eprint{2103.06036}.

\bibitem[{\citenamefont{Di~Marco}(2020)}]{DiMarco:2020mdk}
\bibinfo{author}{\bibfnamefont{N.}~\bibnamefont{Di~Marco}}
  (\bibinfo{collaboration}{GERDA}), \bibinfo{journal}{Nucl. Instrum. Meth. A}
  \textbf{\bibinfo{volume}{958}}, \bibinfo{pages}{162112}
  (\bibinfo{year}{2020}).

\bibitem[{\citenamefont{Agostini et~al.}(2020)}]{GERDA:2017wlm}
\bibinfo{author}{\bibfnamefont{M.}~\bibnamefont{Agostini}} \bibnamefont{et~al.}
  (\bibinfo{collaboration}{GERDA}), \bibinfo{journal}{J. Phys. Conf. Ser.}
  \textbf{\bibinfo{volume}{1342}}, \bibinfo{pages}{012005}
  (\bibinfo{year}{2020}), \eprint{1710.07776}.

\bibitem[{\citenamefont{D'Andrea}(2020)}]{DAndrea:2020ezp}
\bibinfo{author}{\bibfnamefont{V.}~\bibnamefont{D'Andrea}}
  (\bibinfo{collaboration}{GERDA}), \bibinfo{journal}{Nuovo Cim. C}
  \textbf{\bibinfo{volume}{43}}, \bibinfo{pages}{24} (\bibinfo{year}{2020}).

\bibitem[{\citenamefont{Abgrall et~al.}(2021)\citenamefont{Abgrall, Abt,
  Agostini, Alexander, Andreoiu, Araujo, Avignone~III, Bae, Bakalyarov, Balata
  et~al.}}]{abgrall2021legend}
\bibinfo{author}{\bibfnamefont{N.}~\bibnamefont{Abgrall}},
  \bibinfo{author}{\bibfnamefont{I.}~\bibnamefont{Abt}},
  \bibinfo{author}{\bibfnamefont{M.}~\bibnamefont{Agostini}},
  \bibinfo{author}{\bibfnamefont{A.}~\bibnamefont{Alexander}},
  \bibinfo{author}{\bibfnamefont{C.}~\bibnamefont{Andreoiu}},
  \bibinfo{author}{\bibfnamefont{G.}~\bibnamefont{Araujo}},
  \bibinfo{author}{\bibfnamefont{F.}~\bibnamefont{Avignone~III}},
  \bibinfo{author}{\bibfnamefont{W.}~\bibnamefont{Bae}},
  \bibinfo{author}{\bibfnamefont{A.}~\bibnamefont{Bakalyarov}},
  \bibinfo{author}{\bibfnamefont{M.}~\bibnamefont{Balata}},
  \bibnamefont{et~al.}, \bibinfo{journal}{arXiv preprint arXiv:2107.11462}
  (\bibinfo{year}{2021}).

\end{thebibliography}

\end{document}